\documentstyle[11pt,aaspp4]{article}
\begin{document}
\newcommand{\lya}{Lyman~$\alpha$}
\newcommand{\lyb}{Lyman~$\beta$}
\newcommand{\za}{$z_{\rm abs}$}
\newcommand{\ze}{$z_{\rm em}$}
\newcommand{\cmtwo}{cm$^{-2}$}
\newcommand{\nhi}{$N$(H$^0$)}
\newcommand{\nzn}{$N$(Zn$^+$)}
\newcommand{\ncr}{$N$(Cr$^+$)}
\newcommand{\degpoint}{\mbox{$^\circ\mskip-7.0mu.\,$}}
\newcommand{\halpha}{\mbox{H$\alpha$}}
\newcommand{\hbeta}{\mbox{H$\beta$}}
\newcommand{\hgamma}{\mbox{H$\gamma$}}
\newcommand{\kms}{\,km~s$^{-1}$}      
\newcommand{\minpoint}{\mbox{$'\mskip-4.7mu.\mskip0.8mu$}}
\newcommand{\mv}{\mbox{$m_{_V}$}}
\newcommand{\Mv}{\mbox{$M_{_V}$}}
\newcommand{\peryr}{\mbox{$\>\rm yr^{-1}$}}
\newcommand{\secpoint}{\mbox{$''\mskip-7.6mu.\,$}}
\newcommand{\sqdeg}{\mbox{${\rm deg}^2$}}
\newcommand{\squig}{\sim\!\!}
\newcommand{\subsun}{\mbox{$_{\twelvesy\odot}$}}
\newcommand{\et}{{\it et al.}~}

\def\ltsima{$\; \buildrel < \over \sim \;$}
\def\simlt{\lower.5ex\hbox{\ltsima}}
\def\gtsima{$\; \buildrel > \over \sim \;$}
\def\simgt{\lower.5ex\hbox{\gtsima}}
\def\arcs{$''~$}
\def\arcm{$'~$}
\title{QSO ABSORBING GALAXIES AT $z \simlt 1$: DEEP IMAGING AND SPECTROSCOPY IN 
THE FIELD OF 3C~336\altaffilmark{1,2}}
\author{\sc Charles C. Steidel\altaffilmark{3,4,5}}
\affil{Palomar Observatory, Caltech 105--24, Pasadena, CA 91125}
\author{\sc Mark Dickinson\altaffilmark{5}}
\affil{Space Telescope Science Institute, 3700 San Martin Drive, Baltimore, MD
21218}
\author{\sc David M. Meyer\altaffilmark{5}}
\affil{Department of Physics and Astronomy, Northwestern University, Evanston, IL
60208} 
\author{\sc Kurt L. Adelberger}
\affil{Palomar Observatory, Caltech 105--24, Pasadena, CA 91125}
\author{\sc Kenneth R. Sembach\altaffilmark{5}}
\affil{John Hopkins University, Department of Physics and Astronomy, 3400 N.
Charles St., Baltimore, MD 21218}

\altaffiltext{1}{Based in part on observations obtained with the NASA/ESA
Hubble Space Telescope, which is operated by the STScI for the
Association of Universities for Research in Astronomy, Inc., under NASA
contract NAS5-26555.}
\altaffiltext{2}{Based in part on observations obtained at the W.M. Keck
Observatory, which is operated jointly by the California Insitute of
Technology and the University of California.} 
\altaffiltext{3}{Alfred P. Sloan Foundation Fellow}
\altaffiltext{4}{NSF Young Investigator}
\altaffiltext{5}{Visiting Astronomer, Kitt Peak National Observatory}
\begin{abstract}
We present very deep WFPC2 images and FOS spectroscopy from the Hubble Space 
Telescope ({\it HST}) together with numerous 
supporting ground--based observations of the field of the quasar 3C~336 
($z_{em}=0.927$). The observations are designed to investigate the nature 
of galaxies producing metal line absorption systems in the spectrum of the QSO.
Along a single line of sight, we find at least 6 metal line absorption
systems (of which 3 are newly discovered) ranging in redshift from 0.317 
to 0.892.  Through an extensive program of optical and IR imaging, QSO 
spectroscopy, and faint galaxy spectroscopy,
we have identified 5 of the 6 metal line absorption systems with luminous 
($L_K \ge 0.1 L_K^{\ast}$) galaxies.  These have morphologies ranging from 
very late--type spiral to S0, and exhibit a wide range of inclination and 
position angles with respect to the QSO sightline.  The only unidentified 
absorber, despite our intensive search, is a damped Lyman $\alpha$ system 
at $z_{abs}=0.656$.  Analysis of the absorption spectrum suggests that the 
metal abundances ([Fe/H]$=-1.2$) in this system are similar to those in 
damped systems at $z \sim 2$, and to the two other damped systems for which 
abundances have been determined at $z <1$.  The absorption line system must 
either be associated with an underluminous, late--type spiral galaxy which 
we find at a projected disk impact parameter of $\sim 120h^{-1}$ kpc, or with 
an as yet unseen, extremely faint galaxy ($L < 0.05 L_K^{\ast}$) near the QSO 
sightline that eludes detection despite our deep {\it HST} and high 
resolution ground based near--IR images.   We have found no examples of 
intrinsically faint galaxies ($L < 0.1 L^{\ast}$) at small impact parameters 
that might have been missed as absorber candidates in our previous 
ground--based imaging and spectroscopic programs on MgII absorbing galaxies.
We have, however, identified several intrinsically faint galaxies within 
$\sim 50h^{-1}$ kpc of the QSO sightline which do {\it not} produce 
detectable metal line absorption. There are no bright galaxies ($L > 0.1 L_K$) 
within $50h^{-1}$ kpc which do not produce detectable metal lines 
(of Mg~II $\lambda\lambda 2796$, 2803 and/or C~IV $\lambda\lambda 1548$, 1550) 
in the QSO spectrum.  All of these results generally support the inferences 
which we have previously reached from a larger survey for absorption--selected 
galaxies at $z\simlt 1$.

There are several other galaxies with redshifts near that of 3C~336,
suggesting that the QSO is situated in an overdense region, perhaps 
a galaxy cluster.  Previously published reports of a cluster around 
3C~336 were largely misled by the presence of many foreground galaxies 
seen in projection near the QSO.  It is possible that a reported 
measurement of weak shear gravitational lensing in this field
may be produced by the QSO cluster itself, as there appear to be no 
other groups or clusters in the foreground.  We find no evidence for 
a normal, bright QSO host galaxy, although there are several faint 
objects very close to the quasar and at similar redshift which might 
either be companions or part of a disorganized QSO host.  

\end{abstract}

\section{INTRODUCTION}
            
While a relationship between
galaxies and metal--line absorption systems in the spectra of QSOs 
has long been assumed because of a large amount
of indirect evidence based on the statistics, kinematics, and metal content of
the absorbing clouds, it was not until the work of Bergeron \& Boiss\'e (1991)
that the direct connection between apparently normal galaxies and the absorption
systems was explored (although the connection remained controversial --
see, e.g., Yanny \& York 1992). More recently, we have compiled a much larger sample, 
allowing the details of the statistical connection between the absorbing gas
and normal high redshift galaxies to be examined. The aims of
the survey include 
understanding how the gaseous component of the galaxies is 
related to other observable galaxy properties, discriminating which galaxies are potential
absorbers and which are not, and ultimately, using galaxies selected 
by gas cross-section to study galaxy evolution in general 
(Steidel, Dickinson, \& Persson 1994 [SDP]; Steidel 1995; Steidel \& Dickinson 1995).    
The advantages of an absorption--selected galaxy sample for studying galaxy evolution
are many (see SDP), provided
that one can establish the absorber--galaxy connection at redshifts where
galaxy properties can be verified using direct imaging and spectroscopy. 

The sample from which we have begun to draw statistical inferences 
has been based on extensive ground based optical and infrared imaging
of the fields of QSOs comprising surveys for Mg~II $\lambda\lambda 2796$, 2803 doublets
at redshifts $0.25 \le z \le 1.6$ (Steidel \& Sargent 1992), 
with follow--up spectroscopy of candidate absorbing
galaxies.  A general conclusion at $z \simlt 1$ is that the absorbing galaxy population is drawn
from normal, relatively luminous (presumably Hubble--sequence) galaxies whose 
properties do not appear to evolve substantially over the redshift
range observed, and indeed are similar in luminosity and color distribution to the
present--day galaxy population. This lack of strong evolution in the ``normal'' galaxy
population to $z \sim 1$ now appears to be broadly consistent with the results of field galaxy
redshift surveys (Lilly \et 1995) and with morphological studies of faint galaxies in deep images from the Hubble Space Telescope ({\it HST}) 
(Glazebrook \et 1995). 

However, many details concerning the nature of the absorbing material and its relation
to the identified galaxies remain largely unexplored. For example, 
the absence of ``interloper'' galaxies -- that is, galaxies having the same overall
properties and impact parameters as the absorbing galaxies but which produce
no detectable absorption in the QSO spectra -- and the success in identifying a
putative absorbing object in a very high percentage of cases, is remarkably consistent with
a simple model in which every galaxy of luminosity $L_K \simgt 0.1 L_K^{\ast}$ is
a potential absorber. The model assumes a spherical ``halo'' whose extent (for producing
Mg~II $\lambda \lambda 2796$, 2803 absorption of equivalent width greater than
0.3\AA\ in the rest frame) is dependent
weakly on near--IR luminosity (see Steidel \& Dickinson 1995, Steidel 1995).  
By implication, the intrinsically faint, very blue galaxies which appear
to dominate the evolution of the overall field galaxy population at intermediate
redshifts do not contribute significantly to the total gas cross-section,
as they are essentially absent from an ``absorption--selected'' sample. However,
it has been difficult to show this directly simply because intrinsically faint
galaxies are difficult to observe at the typical redshifts of our survey
($\langle z \rangle = 0.65$).  In addition, there are many galaxies in the
sample of absorbers which show little or no evidence for ongoing star formation,
suggesting that substantial near--IR luminosity (which is expected to be more or less
proportional to stellar mass), not current star formation, is the most important
pre-requisite for a galaxy to possess an extended envelope of diffuse gas. By extension,
we infer that early--type galaxies make a substantial contribution to the absorbing
galaxy population. 

The inferences regarding the gaseous geometry are at this point tentative; it is clear that
only improved statistics will allow a definitive statement regarding the
true geometry of the absorbing regions and their relationship to the host galaxies (see, e.g.,
Charlton and Churchill 1996). Nevertheless, the covering fraction of the absorbing
gas must be near unity within $\sim 40h^{-1}$ kpc for a galaxy of luminosity $L_K^{\ast}$, 
and a substantial fraction of the galaxy colors and spectroscopic properties suggest
early morphological types.  We believe that both of these facts argue against extended
disks as the generic origin of the absorbing material. 

In view of these somewhat surprising results, we set out to add a new dimension
to our study of absorbing galaxies at intermediate redshifts by using {\it HST}
images to {\it directly} observe galaxy morphological types and sky orientations in order
to test our inferences about the nature of the gaseous ``halos''. We would also
take advantage of the increased spatial resolution and sensitivity to search
for objects very near the QSO sightline that might have been missed in the
ground--based images upon which all of our conclusions have been based to date. 
An additional motivation for the {\it HST} observations would be to eventually
relate the detailed kinematics of the absorbing gas accessible through high
resolution spectroscopy of the QSOs, and the morphology, orientation, and
environment of the absorbing objects, toward understanding the origins of
the absorbing gaseous envelopes.  

We chose the field  of the quasar 3C~336 as the pilot study for this project
because it is exceptionally rich in moderate redshift absorbers
(Steidel \& Sargent 1992).   Two of the intervening Mg~II systems, at $z_{abs}=0.656$ and
$z_{abs}=0.891$, were considered candidates to be moderate redshift
damped Lyman $\alpha$ systems on the basis of the strong Mg~II and Fe~II lines
and the proximity of what we believed to be the absorbing galaxies. We were interested
in the possibility of using {\it HST} ultraviolet spectroscopy to measure
the H~I column densities in these 2 systems, together with ground--based
searches for Zn~II and Cr~II lines associated with the systems, to
obtain morphology {\it and} chemical abundances/gas--phase depletions for the same
galaxies at moderate redshift. Because the performance of the post--refurbishment 
instrument was uncertain at the time the observations were proposed, we chose
to obtain exceptionally deep data in a single, well--chosen, field rather than
to image a number of fields with exposure times that might turn out to
be insufficient for the scientific objectives. 

The 3C~336 field was the
subject of a previous (ground--based) study during the early stages of our survey
for moderate redshift absorbing galaxies (Steidel \& Dickinson 1992; hereinafter
SD92). With the benefit of not only the {\it HST}
data but also greatly improved ground--based data, we find that there
were a number of incorrect inferences in the earlier work.  
The field of 3C~336 has
also been studied by Bremer \et (1992) in connection with extended emission
at the QSO redshift, and by Hintzen, Romanishin, \& Valdes (1991) and
Rakos \& Schombert (1995) in connection with the possibility of a 
cluster associated with the quasar. 
In the course of studying the absorbing
galaxies in the field, we have also acquired significant new data on the evironment
of 3C~336, which we summarize in \S 4.  
Finally, Fort \et (1996) have suggested that 3C~336 may be an example of a
QSO that has been gravitationally lensed by a foreground group; our extensive galaxy
spectroscopy also allows us to address this possibility.       

In this paper
we present the {\it HST} imaging and spectroscopy of the QSO, together with extensive
new ground--based data, which we use for a comprehensive study of the
of the absorber--galaxy connection along a particulary rich line of sight. 

\section{OBSERVATIONS AND DATA}

\subsection{Optical Imaging}

A total integration of 24,000 s was obtained through filter F702W using the WFPC2 camera aboard the
{\it Hubble Space Telescope} in 1994 September. Each individual exposure
was 1000 s, with 3C~336 centered in the WF2 chip. The integrations
were obtained in 3 slightly offset positions, allowing both very good
cosmic ray filtering (using the sets of 8 registered images) and
bad pixel rejection. 
The 3 stacks of 8 images were put into sub-pixel
registration and averaged to produce the final frame, the directly relevant portion of which
is presented in grayscale in Figure 1 (plate xx).  The total integration time
makes the final image among the deepest obtained thus far with WFPC2, and
was deemed necessary to allow reliable morphological classification of
galaxies as faint as ${\cal R} \approx F702W_{AB} \sim 24$. 

A deep image of the 3C~336 field was obtained in 1995 March at the Michigan--Dartmouth--MIT
2.4m Hiltner telescope through the ${\cal R}$ filter [6930/1500] (a very close
match in response to the {\it HST} filter F702W). The detector was a thinned
Tektronix 1024 X 1024 CCD, with 0\secpoint275 pixels at the f/7.5 Ritchey-Chr\'etian
focus. The total integration time was 6000s, with typical individual exposures
kept to $< 600$s to prevent saturation of the QSO and potential PSF stars
on the CCD. The seeing was reasonably good, resulting in a stellar image size on  
final co--added frame of 0\secpoint95. 
The data were reduced conventionally, using flat fields produced from
a combination of dark--sky (object) frames (suitably masked)  and twilight sky
observations. The photometric calibration was accomplished using spectrophotometric
standards from Massey \et (1988); the ${\cal R}$ magnitudes are on the
AB system, and correspond to an effective wavelength of $6930$ \AA\ (see
Steidel \& Hamilton 1993 for conversions of ${\cal R}$ magnitudes to
standard photometric systems).  
 
This deep ground--based image provides
a much better match to the near--IR image (for the purpose of measuring accurate
optical/IR galaxy colors) than does the {\it HST} image, and 
careful PSF subtraction reliably uncovers all of the objects seen in the
much higher resolution WFPC2 image -- see \S 2.3 below.  
A contour plot of the inner 2\arcm\ region of the field is shown in Figure 2a.

\subsection{Near Infrared Imaging}

Deep near-infrared ($K_s$) band images of the 3C~336 field were obtained in
1993 April using the Kitt Peak 4m Mayall telescope and the IRIM NICMOS 3
256 X 256 camera at the f/15 Cassegrain focus. The plate scale using this
set--up is quite favorable to deep imaging of relatively large fields, with
the 0\secpoint6 pixels providing a total field of view of 2\minpoint5 on a
side.  The total integration
was 2400s, obtained in 120 individual 20s exposures, moving the telescope
over an in--field raster pattern every 60s. The data were reduced
using DIMSUM (Eisenhardt, Dickinson, Stanford \& Ward, private communication).
Each 60s co-addded frame was linearized using a pixel--by--pixel linearization
scheme, flat--fielded using dome flats, and sky subtracted using a temporally local sky,
produced from a masked median of bracketing, dis--registered exposures. 
Each sky--subtracted image was resampled to 4 times the observed
pixelization prior to shifting and adding in an effort to recover
some of the resolution lost because of the spatial undersampling of
the detector. The background
in $K_s$ during the observations ranged from $K=13.4-13.5$ magnitudes per square
arc second, and the final image reaches $K \sim 22$ for the regions of
the mosaic receiving the full exposure time, with stellar image size
of 1\secpoint0. The photometric zero points were established using a number
of stars from the UKIRT Faint Standards list (Casali \& Hawarden 1992). A contour plot of the inner 2\arcm\
region of the image is shown in Figure 2b.

An  additional $K_s$ band image of the central portion of the 3C~336 field
was obtained in 1996 March using the Near Infrared Camera (NIRC) (Matthews
and Soifer 1993) on the
W.M. Keck telescope. The total exposure time was 2160 seconds, with seeing
of $0\secpoint55$ FWHM. The data were reduced using the same procedures 
as for the KPNO IR images. Standards from the Persson (1995, private
communiation) faint standards
list were used to calibrate the data. A contour plot of the NIRC image is shown in figure
2c.   

\subsection{Photometry and PSF Subtraction}

To facilitate the discussion of the galaxies near the line of sight to 3C~336, we
have collected the results of the ${\cal R}$ and $K_s$ photometry for objects within
50\arcs\ of the QSO (the full photometric database is available on request) in Table 1. 
All of the photometry was performed using an enhanced version of FOCAS (Valdes 1982;
Adelberger 1996), in which the images were registered, adjusted to a common point source
FWHM, 
and then the detection phase
was run for each of the images. In order to produce a complete census of all objects
detected, we adopted for each object the maximum isophotal detection area, 
so that
both very blue and very red objects are included in the final catalog. Colors and limits
on colors were obtained through this common optimal aperture; limits on the colors
represent 2 $\sigma$ values based on the local photon statistics. The tabulated ${\cal R}$
and $K$ magnitudes, on the other hand, are FOCAS ``total'' magnitudes, in which the
isophotal area is grown by a factor of 2 in area. Thus, there are a few objects for
which colors are quoted but for which one or the other of the magnitudes is stated
as a limit; this results because of the additional noise introduced in growing the
photometric aperture. The $K$ magnitudes from the Keck image are used in the region
covered by the higher quality and deeper image, and those from the KPNO image are
used beyond the central $\sim 20$\arcs\ radius region.  

Because we are particularly interested in objects very close to the QSO line of sight
(in order to be confident in any particular case that we have identified the most
probable galaxy responsible for an absorption system, and in order to assess
the presence of possible non--absorbing galaxies), we have performed careful
PSF subtraction in both the {\it HST} image and all of the ground--based images. 
For the KPNO $K$ image and the MDM ${\cal R}$ image, we have modeled the PSF
using stars in the same image, scaled this analytic PSF, and subtracted
the QSO light profile. For the Keck $K$ image, a PSF star was observed
immediately following the observation of 3C~336 at the same rotator angle,
sampling the detector in an identical manner. In Figure 2 we present
contour plots of these PSF--subtracted images, where the original
position of the QSO is placed at the origin of the coordinate system. Figure
2a) clearly shows that the object to the north of the QSO called ``G13'' by
SD92 is resolved into at least 2 objects. There is also evidence of a knot
of emission about 1\arcs\ to the south of the QSO, which was not
recognized in the much shallower SD92 data. The objects to the North
are also clearly seen in the $K$ band images. The same objects are
evident in the {\it HST} F702W image (including the knot of emission to the South), 
without PSF subtraction (Figure 1). 
We have
experimented a great deal with the subtraction of the (saturated) PSF in the
WFPC2 image, using both analytic and theoretical PSFs, without much
success in yielding any believable objects closer than $\sim 0\secpoint5$ 
to the QSO line of sight. The great complexity of the {\it HST} PSF, together with
the fact that there is significant saturation in the core of the QSO because
of the length of the individual WFPC2 exposures, yield the surprising
result that the ground--based images are just as effective in allowing
one to find objects very near the QSO sightline as the much higher
resolution WFPC2 images. Nevertheless, the consistency of the observed
structure near the QSO sightline in all of the data is encouraging.
In Figure 2d we show expanded views of the region near the QSO in
the ground based optical and near-IR images and the WFPC2 image, with
the QSO subtracted in each case. We will return to a discussion of the faint objects
very near the QSO sightline in \S 4. In fact, it turns out that 
these objects
all have roughly the same redshift as the QSO, and so no new potential
absorbing objects have been revealed despite the wealth of new and improved
imaging data. 

\subsection {QSO Spectroscopy}

Spectra of 3C~336 were obtained using the {\it Faint Object Spectrograph}
(FOS) on board the {\it Hubble Space Telescope}, with the  
red digicon detector and gratings G190H and G270H and the 0\secpoint9 aperture,
resulting in spectral resolution of $\sim 1.4$\AA\ in the range 1650--2300 
(G190H) and $\sim 2$\AA\ in the range 2200-3300 (G270H). The total integration
time on 3C~336 ($V \sim 17.8$) was 3650s for G270H and 4350s for G190H. 

The pipeline--processed spectra were deemed satisfactory for our purposes, and 
the individual exposures (and associated error vectors) were combined using
standard STSDAS and IRAF tasks. The final spectra, together with their
associated error vectors, are plotted in Figure 2. We noted a slight over-subtraction
of the background at the shortest wavelengths in the G190H spectrum, evident
because the Lyman continuum region of the optically thick Lyman limit 
system (at $z_{abs}=0.891$, to be discussed below) goes slightly below zero intensity.
However, the over-subtraction appears to have been wavelength--dependent, as
no such over-subtraction is evident near the strong Lyman $\alpha$ absorption feature
at $\lambda 2014$. In any case, the effect is very small, and the spectra are
plotted without any zero--point correction to the pipeline--processed data. 

The absorption lines in the FOS spectra were measured using techniques similar
to those used,  e.g., in Steidel \& Sargent (1992). Lines which exceeded 4.5$\sigma$
significance were retained for the linelist, which is given in Table 2. Absorption
systems were then identified first by searching for lines associated with previously--known 
metal line systems at $z_{abs} = 0.3179$, 0.4721, 0.6563, and 0.8912 and Galactic 
absorption features,  and then searching for new metal--line systems. This process
resulted in the line identifications listed in Table 2. The search resulted in the
discovery of a very strong C~IV doublet at $z_{abs}=0.7973$, and subsequent
ground--based spectra of 3C~336 have turned up a weak Mg~II doublet at that redshift
as well (see below). The H~I column density in the new system is not well-constrained
by the observations, since the optically thick Lyman limit at $z=0.892$ prevents
a measurement of the Lyman continuum optical depth; the Lyman $\alpha$ line
is present, but relatively weak ($W_{\lambda}=1.37$ \AA\ rest).  The line
strengths and ionization of the various systems of interest will be discussed in
\S 3.

New ground--based spectra of 3C~336 were obtained during observing runs at the
Lick Observatory 3m Shane telescope in 1993 May, and at the Kitt Peak 4m
Mayall telescope in 1994 June. For both observing runs the aim was to obtain
data covering the expected positions of Zn~II and Cr~II absorption lines
corresponding to the 2 Mg~II systems which were deemed to be potential 
damped Lyman $\alpha$ (DLA) systems on the basis of small galaxy impact
parameters and the unusual strength of the Mg~II $\lambda\lambda 2796$, 2803
doublet.  For this reason, the observations were concentrated in the
spectral region shortward of 4000\AA. The Lick observations used the
Kast Double Spectrograph, for which the blue--side setup had a
Reticon 1200 X 400 CCD and an 830 line/mm grism blazed at 3400 \AA. The 
dispersion was 1.17 \AA\ per pixel, for an effective spectral resolution 
of $\sim 2.5$ \AA and wavelength coverage from 3300--4700. A total of
9 hours of integration was obtained using the Lick 3.0m. In the end, the Lick spectra
were used to corroborate weak features detected in the KPNO spectrum, and they
will not be used further in this paper.

The KPNO observations of 3C~336 were obtained with the RC
spectrograph and a Tek 2048 CCD on the 4m Mayall telescope.  The
spectrograph was configured with the B\&L 420 grating to
cover the 3200 - 4500 \AA\ wavelength region at
a spectral dispersion of 0.76 \AA\ pixel$^{-1}$.  As measured
from the FWHM of the emission lines in our He-Ne-Ar calibration
lamp exposures, the 250 $\mu$m slit width resulted in a
spectral resolution of 1.7 \AA\ over this wavelength region.
Over the course of three nights, a total of 17 exposures spanning
17.25 hr of integration time were taken of 3C~336 at three slightly
different grating tilts centered near 3800 \AA\@.  Using the
IRAF data reduction package, the CCD frames comprising these
observations were bias-corrected, flat-fielded, and
sky-subtracted in the standard manner.  The S/N ratio of the
summed spectrum illustrated in Figure 4a reaches a maximum near 70
in the 3900 - 4300 \AA\ wavelength region.  The absorption lines
detected in this spectrum at a 4$\sigma$ (and greater) confidence
level are identified in Table 3\@.

In addition to the UV spectra described above, additional spectra of 3C~336 were
obtained in the course of performing faint galaxy spectroscopy using the 10m
W. M. Keck telescope and the Low Resolution Imaging Spectrograph (Oke \et 1995) in 1995 May. 
These observations were obtained using a 600 line/mm grating blazed at 7500 \AA\
in first order, which in combination with the 1\secpoint0 slit resulted in
a spectral reslution of 3.5\AA\ and a wavelength coverage of 4900--7450 \AA. 
The total integration time was 2400s at a position angle chosen to include
very close, faint companions to the QSO (see \S 4). 
The seeing was substantially sub--arcsecond  
during this observation, judging from the spatial width of the QSO profile.   
The spectrum is displayed in Fig. 4b, and the measured absorption lines
and identifications are summarized in Table 4.

\subsection{Faint Galaxy Spectroscopy}

Spectra of a number of faint galaxies in the field of 3C~336 were obtained on
the nights of 1 and 2 February 1995 with the 10m W. M. Keck telescope and
the Low Resolution Imaging Spectrograph. The observations were made using
a long slit, oriented at two different position angles on the plane of
the sky. The first observation was made at PA 160 degrees , offset 3\secpoint\
E of 3C~336 to center on the galaxy \#3, with a slit width
of 1\secpoint5. A total integration of 2400s was obtained. The second
observation was made at PA 97 degrees, offset from the QSO to center on 
galaxy \#11, with a slit width of 1\secpoint0 and a total integration time
of 6000s. Both observations used
a 300 line/mm grating, providing wavelength coverage from 3800--8900 \AA\ at
a resolution of $\sim 12$\AA\ and $\sim 8$\AA\ , respectively. 
The data were reduced using standard
procedures; internal quartz lamp flats taken at the same position as the
galaxy observations, together with dome flats used to remove the slit
function on large spatial scales,  were found to result in excellent sky subtraction even
in regions of very complex and strong OH bands in the near--IR.  Some of the 
spectra obtained (a total of 16 faint galaxy redshifts were obtained with only the 2 slit
positions) are plotted in Figure 5. 

As mentioned above, additional spectra in the field were obtained, over a more
restricted wavelength range because of a higher dispersion grating, on the
night of 28 May 1995 (UT). This observation was centered on the QSO itself,
with the slit oriented at PA 165 degrees. The purpose of this observation was to
obtain a spectrum of a ``knot'' located about 1\arcs\ S of the QSO position
that is evident in the WFPC2 images (and the PSF--subtracted ground--based
${\cal R}$ images) and to obtain a high S/N spectrum of the QSO near
5000 \AA\ to search for Mg~II absorption associated with the $z=0.7973$ C~IV
absorption system discovered in the FOS spectrum described above.  

Additional long slit observations of several galaxies near the QSO were obtained
in 1995 September with Keck/LRIS, using the 300 line/mm grating as in 1995 February.

Finally, observations were obtained in 1996 May, again with LRIS, with a slit mask
oriented at PA$=-0.4$ degrees. The total exposure time was 5300s, and again the
300 line/mm grating was used for typical wavelength coverage of 4200-9200 \AA.
These observations were obtained as part of a comprehensive program to obtain
spectra of faint galaxies to ${\cal R} \sim 24.5$ in the fields of QSOs in
our {\it HST} imaging programs, in an effort to establish both the environments
of the absorbers, and to vastly improve our knowledge of the transition 
between absorbing and non--absorbing galaxies. 

At the time of this writing, we have obtained more than 50 redshifts for faint objects within a few
arc minutes of 3C~336, ranging from $z=0.261-1.787$; redshifts are indicated 
for objects within $\sim 50$\arcs\ of the QSO in Fig. 1, and are summarized in
Table 1. Examples of the galaxy spectra are presented in Figure 5.

\section{PROPERTIES OF THE ABSORPTION AND THE ABSORBING GALAXIES}

In order to facilitate the discussion of the galaxies in the 3C~336 field
and the absorption they produce, we have included the redshifts of galaxies within
$\sim 50$\arcs\ of the QSO as presented in Table 1, on the {\it HST} image
presented in Figure 1. The coordinate grids on figures 2a,b,and c can be used to identify
other objects from Table 1 with their corresponding images in figure 1. 
In this section, we discuss each metal line
absorption system, and the galaxy which we believe is responsible
for the absorption, in turn. In Table 5, we have compiled a list
of {\it all} galaxies with measured redshifts smaller than that of the QSO, and
within a projected distance of $< 100 h^{-1}$ kpc, for reference.
In each case, we have converted the the observed ${\cal R}$ and
$K$ magntidues into {\it rest--frame} B and K magnitudes using
a method outlined in SDP, and we  
quote the luminosities at these wavelengths relative to present--day
values of $L^{\ast}_B$ and $L^{\ast}_K$, taking $M^{\ast}_B = -21.0$ (e.g.,
Ellis \et 1996)
and $M^{\ast}_K = -24.9$ (Mobasher \et 1993; Cowie \et 1996) for
$H_0 = 50$ \kms Mpc$^{-1}$.  We also list in Table 5 the equivalent widths,
or limits on equivalent widths, for MgII $\lambda 2796$, C~IV $\lambda 1549$,
and Lyman $\alpha$ at the redshift of each galaxy. 

\subsection{$z_{gal}=0.3181$}

This galaxy is evidently an early type spiral or S0 (\#16). A disk is
evident in the WFPC2 image, but there is no sign of spiral structure.  
The
stellar spectral features are remarkably weak, particularly within the red nuclear
regions of the galaxy. Weak H-$\alpha$, [NII], and [OII] $\lambda 3727$ are
present, but there is no sign of [OIII] or H-$\beta$, or of Balmer absorption
lines (see Fig. 5). Because the spectrum is so devoid of features, the redshift
suggested by SD92 on the basis of much lower quality spectra was incorrect 
by nearly 2000 \kms. The associated Mg~II absorption system is weak and is
found in the very steep wing of the C III] $\lambda 1909$ emission line of the QSO, so that
it was not found in the initial spectra of SS92. The C~IV doublet is quite
strong, on the other hand, and is unambiguous in the FOS spectrum.   

\subsection{$z_{abs}=0.3675$}

The associated Mg~II system was discovered only in the very high quality
KPNO 4m spectrum of the QSO; the blue component of the Mg~II doublet
is blended with the Fe~II $\lambda 2600$ feature at $z_{abs}=0.4717$, but 
we can assume, based on the strength of the other Fe~II features in the
latter system, that the $\lambda 2796$ component of the Mg~II doublet
has a rest equivalent width of $\sim 0.3$ \AA. The FOS spectrum shows
no obvious C~IV feature, but the sensitivity of the spectrum is such
that only lines stronger than $\sim 1$\AA\ observed equivalent width could
have been found with confidence. The galaxy which we suggest is producing the
absorption is a very bright early type spiral, \#45 in Table 1 (see Fig 5.). There is
one other galaxy (\#52) for which we have a spectrum which has the same redshift; it
is a very intrinsically faint galaxy that is also quite red in ${\cal R}-K$ 
with a composite spectrum showing both strong Ca~II lines and
nebular emission.

\subsection{$z_{abs}=0.4721$}

This galaxy (\#4), originally identified in SD92, has a compact morphology with
very high surface brightness.   
It is not obviously a normal spiral, although the
oblong shape might be interpreted as an inclined high surface brightness disk.
It evidently produces C~IV and Mg~II absorption of approximately
equal strength in the QSO spectrum, at an impact parameter of $22h^{-1}$ kpc. 
The absolute magnitude of \#4 places it approximately at the luminosity of
the LMC (see Table 5). 

\subsection{$z_{abs}=0.6563$}

The {\it HST} FOS spectra reveal that the $z = 0.656$ absorber
toward 3C~336 is a damped Ly$\alpha$ system with an H I column
density of 2 x 10$^{20}$ cm$^{-2}$.  Observations of the
redshifted Zn II $\lambda\lambda$2026, 2063 doublet in such
systems can be used effectively as a metallicity indicator
since Zn II is the dominant Zn ion in H I regions, zinc does
not readily deplete into dust grains, and the Zn II doublet is
not heavily saturated (Pettini et al.\ 1994; Meyer \& Roth 1990).
Unfortunately, our KPNO spectra are not sensitive enough to
detect the weak Zn II lines at $z = 0.656$ toward 3C~336.
Our 2$\sigma$ upper limit of 150 m\AA\ on the observed-frame
equivalent width of the Zn II $\lambda$2026 line corresponds
to a Zn II column density of 6 x 10$^{12}$ cm$^{-2}$ and an
upper limit of 60\% solar on the Zn metallicity of the
$z = 0.656$ absorber.  In comparison, Pettini et al.\ (1995)
have found a typical Zn metallicity of 10\% solar in a sample
of 24 damped systems with $z \geq 1.8$ and similar values
have been reported for the few systems probed at lower
redshift (Meyer \& York 1992; Steidel et al.\ 1995; Meyer,
Lanzetta, \& Wolfe 1995).

Although our KPNO observations do not seriously constrain the
Zn metallicity of the $z = 0.656$ absorber toward 3C~336,
they do provide a nice measure of its Fe II column density
and an interesting upper limit on its Mn II abundance.
Along with our 2$\sigma$ upper limit of 80 m\AA\ on the
equivalent width of the weak Fe II $\lambda$2261 line,
the strengths of the five Fe II lines we detect at
$z = 0.656$ provide a good curve-of-growth solution to
the Fe II column density due to the wide $f$-value range
of these transitions (Cardelli \& Savage 1995).  We measure
a line-width parameter of $b = 35 \pm 3$ km s$^{-1}$ and
$N(Fe II) = 4 \pm 1$ x 10$^{14}$ cm$^{-2}$.  Compared to the
solar abundance of Anders \& Grevesse (1989), this measurement
yields a logarithmic gas-phase iron abundance of $[Fe/H] = -1.2$
in the $z = 0.656$ absorber toward 3C~336.  Taken by itself,
this Fe abundance could be a sign of a low-metallicity,
dust-poor medium like that of other damped QSO absorbers or
one in which there has been substantial dust depletion from
a gas of near-solar metallicity like the Galactic ISM.
A strong hint that the former scenario is appropriate in this
case comes from the low Mn abundance in the $z = 0.656$
absorber.  We measure a 2$\sigma$ upper limit of 60 m\AA\
on the equivalent width of the Mn II $\lambda$2577 line
that corresponds to a Mn II column density of 2 x 10$^{12}$ cm$^{-2}$
and a gas-phase manganese abundance of $[Mn/H] \leq -1.5$.
In the Galactic ISM, Mn is appreciably less depleted into
dust grains than Fe with typical gas-phase abundance ratios
of $[Mn/Fe] \geq +0.6$ (Jenkins 1987).  The fact that
$[Mn/Fe] \leq -0.3$ in the $z = 0.656$ absorber suggests
that the dust content of this system is low.  Furthermore,
such a Mn underabundance is a nucleosynthetic signature that
one would expect in a gas cloud with a base metallicity
similar to that of metal-poor stars in the Milky Way.  Among
a sample of Galactic stars with $-0.2 \geq [Fe/H] \geq -2.4$,
Gratton (1989) has found a clear pattern of Mn underabundances
as a function of decreasing stellar metallicity with a value
of $[Mn/Fe] \approx -0.25$ appropriate for $[Fe/H] \approx -1$.
Considering the uncertainties, these numbers are certainly
consistent with the abundance pattern of the $z = 0.656$
absorber toward 3C~336.  In this regard, it is important
to note that Lu et al.\ (1995) and Meyer et al.\ (1995) have
also found recent evidence of $[Mn/Fe]$ underabundances in the
damped systems at $z = 1.7382$ toward HS 1946+7658 and at
$z = 1.3726$ toward 0935+417 respectively.  Clearly, it would
be worthwhile to confirm the apparent low metallicity of the
$z = 0.656$ absorber with more sensitive Zn II observations.

The galaxy responsible for the $z=0.656$ DLA absorption system
has eluded identification despite intensive investigation. One might
expect that an object with such a large H~I column density would
be found within $\sim 15h^{-1}$ kpc of the line of sight to the
QSO if the galaxy is typical of a nearby disk. We can rule
out with a high degree of confidence any galaxy with
a luminosity larger than $\sim 0.1L_K$ at $z=0.656$ within
$\sim 40h^{-1}$ kpc of the line of sight to the QSO. 
The only object in the field with a spectroscopic redshift in
agreement with that of the absorption system is the relatively
faint, late--type spiral (\#26), located $\sim 65h^{-1}$ kpc to the 
NE of the QSO (see Fig. 1).  
While it is impossible to rule out that the gas seen in absorption
is directly associated with galaxy \#26, we regard it as unlikely. If
one were to assume that the absorption arises in an extended disk,
accounting for the estimated disk inclination and position angles 
for galaxy \#26 with respect to the 3C~336 line of sight, it would require
a minimum disk extent of $\sim 120h^{-1}$ kpc 
(at a large H~I column density) for a galaxy with a luminosity of
less than $0.25L^{\ast}$. Of the galaxies within
$\sim 10$\arcs\ of the line of sight without redshifts, only
galaxy \#9 has an ${\cal R} -K$ color such that it could
plausibly have $z=0.656$ (galaxies \#8, \#13, and \#15 are
both very faint and redder than even an unevolved elliptical
would be at $z=0.656$). 
If placed at $z=0.656$, galaxy \#9 would have $L_B \sim 0.04 L^{\ast}$,
similar to that of the SMC, 
but would have an impact parameter to
the QSO of $41h^{-1}$ kpc. 
Moreover, there is no indication 
that galaxy \#26 is involved in an interaction with another galaxy
at the same redshift (the neighbor, galaxy \#23, is likely to
be an elliptical at the redshift of 3C~336 based upon its
morphology and ${\cal R}-K$ color -- see Table 1).  If one or
more additional galaxies at $z=0.656$ were present, tidal forces might be invoked to  
explain large H~I column densities at large galactocentric
radii such as is found locally in, e.g., the M81 group.   

Another possibility is that the absorbing galaxy
is situated directly beneath the QSO on the plane of the sky; however,
even in this case it would have to be underluminous and/or extremely
compact to have escaped detection in both the psf--subtracted ground
based images and in the WFPC2 images (see \S 2). In addition, we have obtained
several very high quality spectra of the QSO in which the slit would
be expected to include any putative absorber directly on the QSO
sightline, in which case one might expect to have detected line emission
at $z=0.656$; no such emission is present on inspection of any of
our recent QSO spectra. We conclude that the galaxy responsible for
the $z_{abs} = 0.656$ damped Lyman $\alpha$ absorption remains
unidentified, and that it is very likely to be a dwarf galaxy in order
to remain undetected despite our efforts. We will return to a discussion
of this system in \S 5.

\subsection{$z_{abs}=0.7973$}

The weak Mg~II absorption system was found only after an {\it a posteriori}
search using the Keck spectrum shown in Figure 4b; the system was first
recognized based on the very strong C~IV doublet in the FOS spectrum
of 3C~336. The galaxy responsible for the absorption is a $\sim L^{\ast}$, apparently
mid--type spiral (\#11 -- see Figs. 1 and 5), at an impact parameter of $47h^{-1}$ kpc. 
We estimate an inclination angle with respect to the line of sight of $\sim 30$ degrees,
so that if the absorption arises in an extended disk rather than in more symmetric halo,
the QSO sightline would be intersecting the disk at a galactocentric radius of 
$\sim 80h^{-1}$ kpc. We note that this absorption system would not have been included
in the homogeneous sample of Mg~II systems of Steidel \& Sargent (1992) because
both components of the doublet do not exceed 0.3 \AA\ equivalent width.  

\subsection{$z_{abs}=0.8912$}

In the case of the $z = 0.891$ absorber toward 3C~336, the
FOS spectra of the Ly$\alpha$ line profile are not 
definitive with regard to the H I column density due
to the possibility of multiple component structure and the
lack of obvious damping wings.  Nevertheless, we can place
an upper limit of 2.5 x 10$^{19}$ cm$^{-2}$ on the value
of $N(H I)$ from the Lyman $\alpha$ line, and a lower limit
of $\sim 8 \times 10^{17}$ cm$^{-2}$ from the optically thick
Lyman limit in the G190H FOS spectrum (Figure 3a).  
The only species in this system
on which the KPNO spectra provide
a serious abundance constraint is Fe II.
The ratio of our 2$\sigma$
upper limit of 60 m\AA\ on the equivalent width of the
Fe II $\lambda$2261 line with the strength of the much
stronger $\lambda$2374 line is essentially the same as the
ratio of their oscillator strengths.  The lack of
saturation implied by this result is indicative of
substantial unresolved component structure in the line
profile and yields a net Fe II column density of
3 $\pm$ 1 x 10$^{14}$ cm$^{-2}$ in the $z = 0.891$
absorber.  Comparing this Fe II column with the limit on H I
leads to a logarithmic gas-phase iron abundance of
$[Fe/H] \geq -0.4$.  Considering this apparently large Fe
abundance as well as the small H I column, it is likely
that much of the Fe II absorption in the $z = 0.891$
system comes from H II regions.  Thus, it would be particularly
interesting to examine high resolution data on the absorption
line profiles arising from this system (see Churchill, Steidel, \& Vogt 1997).

The galaxy responsible for the absorption is a late--type spiral (\#3) seen nearly edge
on at a projected distance of $15h^{-1}$ kpc. Given the geometry,
is seems unlikely that we are seeing absorption due to disk gas, but
if one were to invoke an extended disk rather than a more symmetric
halo of gas, given the inclination and position angle,
the disk would have a minimum extent of $\sim 45h^{-1}$ kpc.  

\subsection{Other Potential Absorbing  Galaxies}

Given how complete our spectroscopy is for galaxies with ${\cal R} < 24.5$ within
10--15\arcs\ of the QSO line of sight, it is important to make an
{\it a posteriori} search for absorption associated with galaxy redshifts
which were not originally evident from the QSO spectra. There are a number 
of galaxies at interesting redshifts and impact parameters falling into
this category, which we detail below.

\subsubsection{$z=0.7016$}

Galaxy \#30 appears both spectroscopically and morphologically to
be a mid--type spiral galaxy, with roughly $L^{\ast}$ 
luminosity, with an impact parameter to the QSO
sightline of 73$h^{-1}$ kpc. It appears to be nearly 
face--on (ellipticity of isophotes $\sim 0.1$). None of the new QSO spectra
presented here covers the expected position of the Mg~II doublet
at the redshift of the galaxy, but examination of the spectrum 
presented in SD92 allows a 5$\sigma$ upper limit on the rest equivalent
width of Mg~II $\lambda 2796$ of $0.26$ \AA. C. Churchill has
kindly searched his Keck/HIRES spectrum of 3C~336 for Mg~II absorption
at the redshift of this galaxy and obtains a tentative
detection of the $\lambda 2796$ line ($\sim 6\sigma$) of 0.04 \AA, at $z_{abs}=0.7029$, plausibly
within the uncertainties in the redshift measurement of the galaxy.  The limit on the
corresponding C~IV line in the FOS spectrum of 3C~336 is $< 0.29$ \AA. 
However, there is an unidentified line in the FOS spectrum at
2069.03 \AA, which could very well be a Lyman $\alpha$ line at
$z_{abs}=0.7020$. If this is the case, the rest frame equivalent
width of the line is $0.48$ \AA. In any case, the Mg~II absorption is certainly
more than 7 times weaker than those considered in statistical samples
of Mg~II absorbers (SS92, SDP). This would be expected given the 
large impact parameter and the overall impact parameter statistics for
Mg~II absorbers in general (SDP, Steidel 1995).  

\subsubsection{$z=0.6352$} 

This galaxy (\#10) is very compact morphologically, very blue in its
optical/IR colors, and predictably has strong nebular line emission (Fig. 5).
Fitting of the isophotes suggests, however, that the light profile
is well-represented by an exponential profile. 
The rest--frame B luminosity of the galaxy is $0.1L_B^{\ast}$ ($q_0=0.05$),
and the impact parameter to the QSO sightline is 42$h^{-1}$ kpc. 
The 5$\sigma$ limit on the rest equivalent width of Mg~II is quite stringent: $< 0.03$
\AA\ (again, courtesy C. Churchill), while the limit on the strength of a C~IV feature in the FOS
spectrum is 0.12 \AA. Again. however, there is an unidentified line in
the FOS spectrum at 1988.76 \AA, which would correspond to Lyman $\alpha$
at $z_{abs}=0.6359$, and a rest equivalent width of $0.47$\AA. A galaxy
of this absolute $K$ luminosity would be expected to have little chance
of producing detectable Mg~II absorption at such a large impact 
parameter, based on the empirical halo model presented in Steidel (1995).
It is precisely this type of blue, intrinsically underluminous object
which appears to be absent from the population of metal--line absorbers. Nevertheless,
the possible presence of weak Lyman $\alpha$ absorption is intriguing.  

\subsubsection{$z=0.5652$}

This object, \#12, is another blue galaxy dominated by nebular line emission, with
a projected separation from the QSO sightline of $41h^{-1}$ kpc. Again the
limits on the strength of the Mg~II absorption are very stringent -- this
is another good example of a ``non--absorbing'' galaxy. Unfortunately,
the FOS spectrum in the vicinity of the Lyman $\alpha$ line from this system
falls in a relatively noisy region, so that a Lyman $\alpha$ line comparable
to that from the $z=0.635$ galaxy would not have been found. 
The object clearly breaks into 2 parts in the WFPC2 image; we are unable
to say definitively whether the redshift applies to one or both of
the components. However, examination of the Keck $K_s$ image and the
WFPC2 image suggests that the two components are of similar ${\cal R}-K$
color. 

There are 3 other galaxies with known redshifts and within $\sim 85-95h^{-1}$ kpc
of the line of sight (see Table 5). None has any detected absorption of
Mg~II, C~IV, or Lyman $\alpha$, with the possible exception of \#41, for
which there is a possible Lyman $\alpha$ line which falls in a very crowded region
of the FOS spectrum. We regard the identification as quite uncertain, as the
velocity agreement is not particularly good.

A histogram of the measured redshifts in the 3C~336 field is plotted
in Figure 6, and the redshifts of the six strong metal line absorbtion
systems are indicated there as well.  The absorbers appear to be randomly 
drawn from the distribution of galaxies foreground to the QSO, supporting
our assertion (SDP) that MgII absorption is a characteristic of ordinary,
typical field galaxies.   To the degree that our sparse sample is 
representative of the field galaxy distribution, two of the galaxies 
identified as MgII absorbers have companions at similar redshifts elsewhere
in the field, while three occur alone.   (The identification of the 
$z=0.6563$ damped Lyman-$\alpha$ absorber is problematic, as discussed 
in \S3.4.)   Larger redshift samples will be needed to better establish the 
environments of the MgII absorbers, e.g. their relation to galaxy groups, 
clusters, and the narrow ``walls'' seen in distant field galaxy redshift 
distributions (cf. Cohen \et 1996).

\section{THE ENVIRONMENT OF 3C~336}

\subsection{Cluster companions to the QSO}

3C~336 is among the special class of steep spectrum radio quasars, and 
as such its environment has been the subject of several previous studies 
(Bremer \et 1993, Rakos \& Schombert 1995, Hintzen \et 1991).  The latter 
two papers have discussed the existence of a cluster surrounding the quasar 
on the basis of a high surface density of galaxies near the QSO on the plane 
of the sky, and the first paper has cited the presence of extended line 
emission in the vicinity of the QSO as evidence for a large cluster cooling 
flow.

Most of the galaxies which were to have comprised the putative clusters in 
the above--referenced papers are in fact at different redshifts than that of 
the QSO.   E.g., the dense clump to the north of the QSO, which includes 
the prominent S0 galaxy \#16, does indeed contain at least 2 galaxies at 
$z=0.920$, but only one of these is bright enough to have been detected by 
the previous studies.  The other galaxies in this clump for which we have 
spectra have at least 6 different redshifts, all with $z << z_{\rm QSO}$.

Although previous imaging studies were misled to believe that brighter 
foreground galaxies were part of a cluster containing the QSO, our 
spectroscopy provides some evidence for a cluster nevertheless.
The redshifts listed in Table 1 include eight objects in the redshift range 
$0.919 < z < 0.931$:  seven distinct galaxies plus the quasar itself.  
In addition, there are two other continuum ``blobs'' located very close 
to the quasar (1a and 1b in Figure~7 -- see \S2.3 and below, \S4.2) with
concordant redshifts -- these may be additional galaxies or part of the QSO
host.  In the distribution of redshifts shown in Figure~6, the most 
prominent feature is a pronounced spike near the QSO redshift.

Three of the galaxies at the QSO redshift are evidently ellipticals 
(numbers 14, 19 and 64) and have very red colors (${\cal R}-K \approx 5.1$) 
as expected.  There are several other galaxies with similar ${\cal R}-K$ 
colors which also might plausibly be at the QSO redshift, but which currently
lack spectroscopic measurements.  However, in the $R-K$ vs. $K$ 
color--magnitude diagram, no prominent feature is seen that might suggest 
the presence of a very rich cluster (cf. the $z=1.2$ cluster around 
the radio galaxy 3C~324, Dickinson 1996).   While we do not regard it 
as secure that there is a rich cluster associated with 3C~336, the quasar 
is evidently associated with some significant feature in the large--scale 
structure along this line of sight.  

In a separate imaging study aimed at investigating the effects of 
gravitational amplication of background quasars by foreground groups, 
Fort \et al (1996) claimed a detection of weak shear gravitational lensing 
in the 3C~336 field.   Their shear field is apparently centered the clump
of galaxies to the north of the QSO.\footnote{We note that Figure 3 of 
Fort \et has stated the orientation of the field incorrectly. On their image, 
North is to the right, and East is up.}  They refer to galaxy \#16, 
the $\sim L^{\ast}$ S0 at $z=0.318$, as the ``bright central elliptical'' of a foreground
group responsible for the lensing shear, and they infer that this group
may be amplifying the QSO luminosity as well.  As already noted above, 
this ``foreground group'' is illusory and consists of galaxies at many 
different redshifts.  We have found no other galaxies which share the redshift 
of galaxy \#16.  Indeed, Figure~6 shows clearly that there is no prominent 
foreground group at any redshift along this line of sight.  We are unable to 
comment conclusively about the origin of the reported lensing shear.  
Because gravitational lensing depends on the mass {\it surface density} 
along a line of sight, it may indeed be possible that even a purely 
{\it apparent} group of galaxies seen in chance projection could produce 
lensing effects.  However, it is also possible that the shear reported by 
Fort \et might be induced by the QSO cluster itself.  Galaxy \#14 in the 
northern clump is in fact a $\sim 4 L_K^{\ast}$ giant elliptical at $z=0.92$
which may well be the dominant galaxy of the QSO cluster (see \S4.2
below).  While quite faint in the V-band image of Fort \et , galaxy \#14 
is dominant in our K--band image (see Fig. 2c), and is also near the center 
of the shear contours presented by Fort \et.  

To summarize, our spectroscopic data emphasize the danger of making conclusions 
about the presence of very distant clusters on the basis of imaging data 
alone.  In this case, the projection of objects at many different redshifts 
within a fairly small solid angle makes the field particularly problematic 
in this respect.  (The same effect is the reason for the unusually large 
number of metal line absorption sytems along a single line of sight). 
 
\subsection{The QSO host galaxy}

As we have shown in \S2.3 above, there are extensions from the QSO to both 
the North and the South.  These appear to be quite compact on the {\it HST} images, 
and the (at least) two objects to the North (collectively referred to as 
``G13'' in SD92) are clearly detected in the $K$ band. Given our primary aim, 
to establish the nature of the absorbing galaxies along the QSO line of sight, 
it was very important to determine redshifts for these nearby objects, 
particularly in light of the fact that we had yet to identify a plausible 
candidate for the galaxy giving rise to the damped Lyman $\alpha$ absorption 
system at $z=0.656$.  However, we have shown through Keck spectroscopy under 
very good seeing conditions that these objects are all at redshifts within 
$\sim 1000$ \kms of the QSO redshift.  There is also clearly a great deal 
of extended, forbidden line emission in the vicinity of the QSO, some of 
which has associated continuum (e.g., the two objects to the North of 
the QSO -- labeled 1b and 2 on Figures 1 and 7), and some of which 
does not.  In figures 7 and 8, we have show diagrams of the slit positions 
and contour plots of the two--dimensional spectrograms in the vicinity of 
the [OII] 3727 emission line near $z=0.927$.  The line--emitting ``blobs''
exhibit a velocity spread in excess of 1000 \kms.  At the same time, there 
is no evidence from the spectra of emission from any hidden object at 
$z=0.656$ that might be responsible for the damped Lyman $\alpha$ system at 
that redshift.

There is no evidence for a normal, luminous host galaxy to the QSO from our 
imaging data.  As can be clearly seen in the three PSF--subtracted images in 
Figure 2d, there is no sign of any symmetric residual after subtraction of 
the QSO point source, even in the {\it HST} and Keck/NIRC images, which are both 
extremely deep and have excellent image quality. This is somewhat surprising, 
given that radio--loud QSOs are believed to be situated within giant elliptical 
host galaxies (e.g., Boroson, Oke \& Green 1982; Smith \et 1986, 
McLeod \& Rieke 1995).  In fact, galaxy \#14, about 12\arcs\ to the north 
of the QSO, is a giant elliptical at $z=0.921$ with a rest--frame K luminosity 
of about $4L_K^{\ast}$. If such a galaxy were placed precisely ``underneath'' 
the QSO image, it would have been exceedingly obvious in our PSF subtractions.
However, we cannot rule out a much fainter host galaxy with luminosity 
comparable to that of the close companions (1a, 1b and 2 in Figure~7).
It may also be that some of these ``companions'' actually represent pieces
of an extended, chaotic host galaxy, similar to (but substantially 
fainter than) the highly disturbed 3CR radio galaxies at $z \approx 1$ 
seen in {\it HST} images (e.g., Longair \et 1995, Dickinson \et 1995, Best \et 1996.)
Infrared images of these radio galaxies, however, generally show symmetric and 
very luminous host galaxies which would have been easily seen in our NIRC 
data on 3C~336.

%

If there is a cluster associated with 3C~336, then it is curious that
that the QSO itself does {\it not} seem to reside in the dominant cluster
galaxy.  Also, we note that with the exception of the very close quasar 
``companions'' 1a, 1b and 2, all of the other potential cluster galaxies 
lie at redshifts $\sim 1000$ \kms {\it smaller} than that of the QSO 
(the QSO redshift is determined from its forbidden lines).

\section{DISCUSSION}

While the sample of absorbing galaxies we have compiled in this single QSO sightline
is far too small to allow general conclusions about the overall nature of
metal line absorbers in QSO spectra, it can be used effectively in conjunction with
our large ground--based absorbing galaxy survey. Of the six known metal line
absorption system along the 3C~336 line of sight, we have identified the galaxies
we believe to be responsible for the absorption in five of those six cases. As
we had concluded on the basis of our previous ground--based work, the bulk
of the absorbers are indeed relatively luminous galaxies spanning 
the normal Hubble sequence of morphologies. In this small sample, we have
two very luminous early type spiral galaxies,  two mid--type
spiral galaxies (one of which is seen nearly edge--on),
and a very compact, high surface brightness object of relatively low
luminosity which defies simple morphological classification. Despite the much
deeper, higher resolution images we have presented here, as well as 
spectroscopy which reaches fainter magnitudes and achieves greater 
completeness than our previous efforts,
the conclusion that the galaxies reponsible for the absorption
are luminous objects within $\sim 40-50h^{-1}$ kpc remains. There are
also no luminous ($L \simgt 0.1 L^{\ast}$) galaxies within such projected distances which
do not give rise to measureable absorption lines of either Mg~II or C~IV
or both, supporting our previous assertion that the covering fraction for the
extended gas associated with such galaxies must be near unity.  Looking at Table 5,
there is a general, albeit not highly statistically significant (made more significant
if it is assumed that the unidentified $z=0.656$ absorber is very close to the QSO line of sight), trend for the
ratio of C~IV to Mg~II absorption line strength to increase with increasing 
galactocentric impact parameter. Such a trend would be expected if, for
example, the gas density in halo clouds (or the density of gas in an extended
disk) decreases with galactocentric distance, resulting in a higher level
of ionization. Clearly more data would be required to establish this securely,
but the trend is in the expected sense if the gas is physically ``attached'' to the
identified galaxies.  

On the other hand, we draw attention to the fact that the one galaxy which
has not been identified in this field is the one for which we have
the most information from the absorption line spectrum of the QSO:
the $z=0.656$ damped Lyman $\alpha$ absorption system. An analysis
of the absorption line spectrum suggests that the gas in this system
is underabundant in heavy elements by a factor of $\simgt 10$ relative
to solar (\S 3.4), similar to other damped systems at comparable
redshifts (Meyer \& York 1992, Steidel \et 1995). As in these
other two systems with metal abundance measurements, however, 
the galaxy responsible is not a normal, relatively luminous spiral 
galaxy.  Whereas in these other two cases the galaxies were
of low surface brightness and low luminosity, respectively, here
we find no plausible absorbing galaxy at all, after a much more 
intensive search than in any previous line of sight. 
LeBrun \et (1996b) have recently presented results from a program of {\it HST} 
imaging aimed at identifying and studying the galaxies responsible for 
seven confirmed or suspected damped Lyman~$\alpha$ absorption systems 
at $z < 1$.  They find that the absorbing galaxies exhibit a much 
broader range of morphological properties and luminosity than do the 
MgII systems.   While some are readily identifiable with normal disk 
galaxies, others are found to have very low luminosities, low surface 
brightnesses, or extremely compact morphologies.  LeBrun \et identified 
plausible candidates for all of their absorbers, although several do not
yet have spectroscopically confirmed redshifts.
Our inability to find the $z=0.656$ absorbing galaxy implies that either 
(1) the object is very faint and aligned {\it very} closely with the 
QSO line of sight, or (2) the absorption arises from gas with little 
or no associated starlight, perhaps material tidally drawn (by an as 
yet unidentified companion) from galaxy \#26, located $65 h^{-1}$~kpc 
away. 

While
the statistics on low redshift damped Lyman alpha absorbing galaxies
are still quite sparse, the existing data suggest that selecting
galaxies at higher N(H~I) contours significantly changes the
nature of the objects producing the absorption. That is, if 
damped systems identified thus far at $0 < z < 1$ represent a fair sample, then relatively
luminous, Hubble sequence galaxies do not necessarily dominate the total cross-section
of high column density (N(HI)$\simgt 10^{20}$) gas. This is somewhat at odds
with the work of Rao \& Briggs (1994), who conclude that the 
total H~I cross-section in the local Universe is dominated by bright spiral
galaxies; however, there is the possibility of selection effects at work
in this high column density regime because of the possible presence of substantial dust in
chemically evolved galaxies, such that they might be missed in 
surveys dependent on optically selected background QSOs (Pei \& Fall 1995,
Steidel \et 1994). If this were the case, it would naturally explain why
one has yet to find a case of a damped Lyman $\alpha$ system at $z<1$ with
anywhere near solar abundance of metals, and why the subset of the
Mg~II absorbing galaxies with high H~I column densities appears to
be significantly different from the overall sample at large.  

As for the issue of the geometry of extended gaseous regions relative
to the identified galaxies, the {\it HST} sample presented here again
is too small to definitively rule out
either relatively spherical gaseous halos or extended disks. 
We are in the process of compiling a large sample of absorbing
galaxies observed with {\it HST} in which we will present a
much more comprehensive statistical analysis. However, we note
that at at least two of the absorbing galaxies (numbers 16 and 45)
show no spectroscopic evidence for ongoing star formation, and are
of relatively early morphological type (\#16 appears to be an S0),
and several would require enormous disk extents were the absorption
arising in extended thin disks, after accounting for the 
galaxy inclinations relative to the line of sight to the QSO. 
 
It is interesting to note that, while our FOS spectrum of 3C~336
does not have the high S/N of most of the spectra obtained for
the {\it HST} Absorption Line Key Project survey (e.g., Bahcall
\et 1993), we are able to identify essentially every significant
absorption feature in the spectrum with Galactic interstellar absorption, 
a known metal line absorption system, or a Lyman $\alpha$ line plausibly close in 
redshift to one of the galaxies for which we have obtained a redshift.
There are two reasonably secure identifications of ``Lyman $\alpha$ only''
systems with galaxies within $\sim 100h^{-1}$ kpc,  
one at $z_{abs} = 0.635$ and one at $z_{abs}=0.702$. Each of these 
has a rest equivalent width of $\sim 0.5$ \AA, but they are apparently produced by 
very different types of galaxies -- one is a giant spiral and one is a star-forming
dwarf galaxy. The absorption systems are typical
of those that have been the subject of recent studies by Lanzetta \et (1995),
Le Brun \et (1996a), and Bowen \et (1996) in the context of identifying
galaxies associated with low redshift, relatively strong Lyman $\alpha$ lines. While
Lanzetta \et (1994) have concluded that the absorption is directly associated
with luminous galaxies at distances as large as $\sim 160h^{-1}$ kpc, the
latter two papers appear to support a picture in which the low column density
absorbing gas is likely to trace regions containing galaxies, but is not necessarily
directly associated with a single identified galaxy. We merely point out here that
most surveys for moderate redshift Lyman $\alpha$ absorbers would not have
seen the dwarf galaxy (\#10) at an impact parameter of $\sim 40h^{-1}$ kpc, and might
have concluded, therefore, that the absorption has no associated galaxy
or that it is associated with a much brighter galaxy at much larger impact
parameter. Also, given the recent results from field redshift surveys
with the Keck telescope (Cohen \et 1996) that moderate redshift galaxies tend to be found as parts
of ``sheets'' or ``walls'' with very small line-of-sight velocity dispersions
on transverse scales of hundreds of kpc, it is clearly dangerous to 
jump to conclusions about the nature of the very tenuous gas producing the
``Lyman $\alpha$ only'' systems and the association with individual galaxies when
the apparent impact parameters become large. 

\section{SUMMARY AND CONCLUSIONS}

We have obtained the most comprehensive data to date on any single QSO field
in an effort to test hypotheses based on a large ground--based survey
for galaxies giving rise to Mg~II $\lambda\lambda$ 2796, 2803 absorption systems.  
These data include {\it HST} FOS spectroscopy of the QSO, a number of new,
high S/N spectra of the QSO itself, very deep WFPC2 images,
new ground--based optical and near--IR images, and extensive
spectroscopy of faint galaxies in the field of the $z=0.927$ quasar
3C~336.  There are now at least 6 known intervening metal line
absorption systems ranging in redshift from $z_{abs}=0.318$ to
$z_{abs}=0.892$ (with a seventh system probably having extremely weak
metal lines), exhibiting a wide range of Mg~II doublet strength
and ratio of low to high ionization species. We have obtained
the redshifts for more than 30 galaxies within 50 arc seconds
of the line of sight, allowing us to identify the probable absorbing object in
5 of the 6 cases. The identified objects have morphologies of
normal Hubble sequence galaxies (with the possible exception of 
one object, which may be classified as a ``compact narrow emission line''
galaxy) with a variety of impact parameters and inclination angles with
respect to the line of sight to the QSO. The Mg~II absorbers range in
luminosity from $\sim 0.1 L_K^{\ast}$ to $\sim 1.8L_K^{\ast}$, and have
projected impact parameters relative to the QSO line of sight ranging
from $15-74h^{-1}$ kpc. All of the systems which would have been
included in previous homogeneous samples of Mg~II absorption line
systems (both components of the Mg~II doublet exceeding 0.3 \AA\ in rest
equivalent width) have impact parameters smaller than $40h^{-1}$ kpc, in
agreement with our previous work (SDP). 

Our spectroscopic follow--up in this field has reached considerably deeper
than any previous study, with spectroscopic identifications reaching
to magnitude levels as faint as ${\cal R}=24.5$ and $K\approx 22$. We do not
find any intrinsically very faint potential absorbers closer to the
sightline than the galaxies which would have been flagged as the most
likely absorbers in a typical field
from our large ground--based survey. We also do not find any galaxies brighter
than $\sim 0.1L_K^{\ast}$ at impact parameters smaller than $40h^{-1}$ kpc
that do not produce detectable metal line absorption in the QSO spectrum.
We do, however, find two intrinsically very faint ($\sim 0.02L^{\ast}$)
and blue galaxies at impact parameters of $\sim 40h^{-1}$ kpc which
do not produce Mg~II absorption to 5$\sigma$ equivalent width limits of 
$\sim 0.03$ \AA (but one of them appear to produce a significant Lyman $\alpha$
absorption line in the FOS spectrum of the QSO). 
Both of these facts support the hypothesis that
{\it all} field galaxies (apparently independent of morphological type)
with luminosities $L_K \simgt 0.1L_K^{\ast}$
are potential absorbers, but that very intrinsically faint galaxies
do not contribute significantly to the Mg~II absorption cross-section.  
There is a suggestion from the data presented here that extending 
the sensitivity of absorption line surveys to weaker equivalent width
limits than the currently available surveys would result in larger
impact parameters to the absorbing galaxies, and that the ratio
of C~IV/Mg~II absorption line strengths tends to increase with
increasing galactocentric distance. 

The object responsible for the damped Lyman $\alpha$ system at $z_{abs}=0.656$ 
(N(H~I)$=2 \times 10^{20}$ cm$^{-2}$)
remains unidentified, despite our intensive efforts.  
There is a galaxy which has the same redshift as the absorber within reasonable
accuracy, but it is an intrinsically faint ($L_K = 0.07L_K^{\ast}$) late type
spiral with a projected disk impact parameter of more than $100h^{-1}$ kpc.
We are forced to conclude that the object reponsible for the damped absorption system
is beyond the sensitivity of our current survey, requiring that it be
extremely underluminous ($L < 0.05L_K^{\ast}$) and/or less than 0\secpoint5 from
the QSO line of sight. Interestingly, the metal abundances in this system are quite low,
entirely consistent with the handful of other determinations at $z < 1$ and
with the typical abundances at $z \sim 2$.

We have obtained spectroscopic evidence for a cluster of galaxies at or near the redshift
of the QSO, $z_{cl} \approx 0.923$. While the presence of a cluster has been suggested
by several previous studies, it turns out that the high density of galaxies near
the QSO on the plane of the sky used to establish the presence of the cluster
is largely due to the projection of a number foreground
redshifts, including many of the absorbing galaxies. The bona fide cluster galaxies
are generally too faint to have been recognized in previous optical imaging programs, but
they become quite prominent in the near--IR (e.g., K band).  It is possible that
the QSO cluster has induced a significant distortion of faint background galaxies
observed by Fort \et (1996). The quasar itself does not appear to have a luminous    
early--type host galaxy, but there are a number of objects exhibiting a velocity
range of more than 1000 \kms within a few arc seconds of the QSO. The structures
are most prominent in forbidden line emission, although some have significant near--IR
continuum as well. 

We regard this paper as an the first installment of a much larger project which includes
WFPC2 imaging of 24 QSO fields ({\it HST} Cycles 5 and 6 -- see Dickinson \&
Steidel 1996 for a preliminary report) from our absorbing galaxy survey,
and accompanying faint galaxy spectroscopy using the Keck telescopes. As such, we refrain
from drawing too many conclusions on the basis of a handful of absorbing galaxies. However,
we do not find it necessary at present to alter the working hypotheses developed
from our more extensive ground--based survey, despite the
much greater detail afforded by the {\it HST} images.

We would like to thank the staffs of the Michigan--Dartmouth--MIT, Kitt Peak, Lick,
and Keck Observatories for their expert help in this observing--intensive program. We are grateful to Chris Churchill for communicating some results
prior to publication.
We acknowledge financial support from the U.S. National Science Foundation through
grant No. AST-9457446 for much of the ground--based observing, and from
NASA through grant G0-05304.03-93A from the Space Telescope Science Institute,
which is operated by the Association of Universities for Research in Astronomy, Inc.,
under NASA contract NAS5-26555.

\newpage

\begin{planotable}{rrrrrrrrr}
\footnotesize
\tablewidth{0pc}
\tablecaption{Objects in the Field of 3C~336\tablenotemark{a}}
\tablehead{
\colhead{ID No.} & \colhead{$\Delta \alpha$} & \colhead{$\Delta \delta$} 
& \colhead{$\Delta \theta$} & \colhead{${\cal R}$} & \colhead{$K$} &
\colhead{$({\cal R}-K)$} & \colhead{Redshift}  \nl
\colhead{} & \colhead{(\arcs)} & \colhead{(\arcs)} & \colhead{(\arcs)} & & &
&
&
 }
\startdata
1 & 0.00 & 0.00 & 0.00 & 17.95 & 15.53 & 2.44 & 0.927 \nl
2 & $-$1.49 & 2.10 & 2.58 & 24.43 & 21.32 & 3.13 & 0.931 \nl
3 & 2.95 & $-$0.03 & 2.95 & 23.36 & 19.57 & 3.81 & 0.892 \nl
4 & $-$4.15 & $-$3.77 & 5.61 & 22.83 & 19.66 & 3.20 & 0.472 \nl
5 & 7.63 & $-$2.54 & 8.04 & 25.38 & 20.14 & 5.53 &  \nl
6 & 6.14 & 5.91 & 8.52 & 22.25 & 20.37 & 2.07 & star \nl
7 & 8.61 & 0.99 & 8.67 & 23.30 & 19.27 & 3.89 & star \nl
8 & $-$3.18 & $-$8.36 & 8.95 & 24.46 & 19.78 & 4.75 &  \nl
9 & 6.16 & $-$6.52 & 8.97 & 25.60 & 21.56 & 3.99 &  \nl
10 & $-$5.54 & 7.54 & 9.35 & 24.21 & 21.91 & 2.30 & 0.635 \nl
11 & $-$8.93 & 3.20 & 9.49 & 22.88 & 18.59 & 4.15 & 0.798 \nl
12 & $-$3.33 & 9.07 & 9.66 & 24.32 & 21.28 & 3.21 & 0.565 \nl
13 & $-$7.39 & $-$8.35 & 11.15 & $>$26.24 & 21.98 & $>$4.28 &  \nl
14 & $-$1.32 & 11.29 & 11.37 & 22.22 & 17.23 & 5.26 & 0.921 \nl
15 & $-$9.53 & 6.28 & 11.41 & $>$26.11 & 21.84 & $>$4.58 &  \nl
16 & $-$6.69 & 9.74 & 11.82 & 20.20 & 16.22 & 4.24 & 0.318 \nl
17 & $-$11.14 & $-$4.28 & 11.94 & 24.09 & 20.60 & 3.30 & 1.037 \nl
18 & 6.18 & $-$10.48 & 12.17 & 25.66 & 21.60 & 4.20 &  \nl
19 & $-$5.08 & 11.10 & 12.21 & 23.61 & 18.74 & 5.12 & 0.920 \nl
20 & $-$1.10 & $-$12.22 & 12.27 & $>$25.68 & 21.18 & $>$4.46 &  \nl
21 & $-$3.10 & 11.96 & 12.35 & 26.15 & 21.81 & 4.60 &  \nl
22 & $-$10.16 & $-$9.51 & 13.92 & $>$25.73 & 21.16 & $>$4.88 &  \nl
23 & 3.63 & 13.58 & 14.06 & 23.25 & 18.47 & 5.08 &  \nl
24 & 7.31 & $-$12.09 & 14.13 & 22.66 & 18.18 & 4.63 & 0.920 \nl
25 & $-$4.06 & 13.67 & 14.26 & 23.29 & 18.65 & 4.82 &  \nl
26 & 1.61 & 14.19 & 14.29 & 23.55 & 20.77 & 3.07 & 0.656 \nl
27 & $-$6.02 & $-$13.49 & 14.78 & $>$26.36 & 21.91 & $>$4.35 &  \nl
28 & $-$8.73 & 12.74 & 15.44 & 25.59 & 20.95 & 4.74 &  \nl
29 & 11.44 & 10.39 & 15.45 & $>$25.27 & 19.77 & $>$5.96 &  \nl
30 & $-$12.28 & 9.86 & 15.75 & 21.89 & 18.10 & 3.94 & 0.702 \nl
31 & 15.05 & 5.74 & 16.11 & 24.79 & 19.46 & 5.24 &  \nl
32 & $-$15.77 & $-$5.30 & 16.63 & 22.75 & 20.04 & 3.00 &  \nl
33 & $-$15.13 & $-$6.98 & 16.66 & 23.82 & 18.97 & 5.14 &  \nl
34 & $-$13.00 & $-$10.89 & 16.96 & 25.36 & 20.29 & 5.21 &  \nl
35 & 0.58 & $-$17.14 & 17.15 & $>$25.67 & 21.45 & $>$4.46 &  \nl
36 & $-$14.80 & 10.18 & 17.96 & 23.82 & 20.35 & 3.62 &  \nl
37 & $-$6.63 & $-$16.97 & 18.22 & 25.47 & $>$20.90 & $<$4.21 &  \nl
38 & 18.24 & 0.17 & 18.24 & 25.19 & 20.57 & 4.61 & 1.010 \nl
39 & 5.62 & 17.36 & 18.25 & 25.45 & 19.93 & 5.67 &  \nl
40 & 8.23 & 16.79 & 18.70 & 23.87 & 19.32 & 4.70 &  \nl
41 & 7.90 & $-$17.01 & 18.75 & 24.17 & 19.31 & 5.02 & 0.828 \nl
42 & 19.11 & 3.66 & 19.46 & 25.51 & 21.73 & $<$3.62 &  \nl
43 & 11.72 & $-$18.21 & 21.66 & 24.11 & 19.32 & 5.23 &  \nl
44 & 16.73 & 14.59 & 22.20 & $>$25.61 & 20.78 & $>$5.25 &  \nl
45 & $-$21.33 & $-$6.17 & 22.21 & 19.76 & 15.90 & 3.90 & 0.368 \nl
46 & 22.47 & 0.13 & 22.47 & 24.42 & 20.45 & 3.97 &  \nl
47 & 8.53 & 21.37 & 23.01 & 25.38 & $>$20.58 & $<$4.40 &  \nl
48 & 12.04 & $-$19.79 & 23.17 & 23.70 & 20.49 & 3.66 &  \nl
49 & 22.58 & 7.06 & 23.66 & $>$25.59 & 20.61 & $>$5.33 &  \nl
50 & 23.20 & $-$4.72 & 23.67 & 23.38 & $>$20.47 & $<$3.60 &  \nl
51 & $-$13.42 & $-$20.21 & 24.26 & 25.28 & $>$20.57 & $<$4.52 &  \nl
52 & $-$24.07 & 5.27 & 24.64 & 23.58 & 19.52 & 4.05 & 0.368 \nl
53 & 15.01 & 19.59 & 24.68 & 24.01 & $>$20.22 & 3.57 & 1.018 \nl
54 & 1.76 & $-$24.65 & 24.71 & 24.68 & $>$20.88 & $<$4.08 &  \nl
55 & 24.73 & 1.66 & 24.79 & 24.50 & $>$20.74 & $<$3.70 &  \nl
56 & 17.78 & $-$17.44 & 24.90 & 24.96 & 19.64 & 4.96 &  \nl
57 & 9.73 & $-$23.09 & 25.06 & 24.81 & 21.25 & $<$3.77 &  \nl
58 & 22.67 & 11.39 & 25.37 & 25.58 & 21.39 & 4.33 &  \nl
59 & $-$25.41 & $-$3.79 & 25.69 & 25.82 & 20.87 & 5.12 &  \nl
60 & 11.41 & $-$23.32 & 25.97 & 24.80 & $>$21.11 & $<$3.73 &  \nl
61 & $-$4.36 & $-$25.91 & 26.28 & 24.64 & $>$20.22 & $<$4.13 &  \nl
62 & 8.80 & $-$24.84 & 26.35 & 22.22 & 19.86 & $<$2.25 & 0.919 \nl
63 & 2.15 & $-$26.40 & 26.49 & 25.19 & $>$21.20 & $<$4.51 &  \nl
64 & $-$26.38 & $-$2.72 & 26.52 & 22.64 & 17.67 & 5.04 & 0.924 \nl
65 & 0.45 & 26.61 & 26.61 & 21.24 & 17.45 & 3.91 &  \nl
66 & $-$21.32 & 16.79 & 27.14 & 25.41 & $>$20.57 & $<$4.61 &  \nl
67 & $-$25.99 & $-$8.59 & 27.37 & $>$25.65 & 20.14 & 5.44 &  \nl
68 & $-$17.98 & 21.07 & 27.70 & 24.13 & 19.84 & 4.30 &  \nl
69 & 13.12 & 24.78 & 28.04 & 24.87 & $>$20.61 & $<$3.97 &  \nl
70 & 6.46 & 27.43 & 28.18 & 22.28 & 19.24 & 3.40 &  \nl
71 & 26.47 & $-$10.00 & 28.30 & 21.34 & 17.90 & 3.42 &  \nl
72 & 25.39 & 13.46 & 28.73 & 24.50 & 19.08 & 5.47 &  \nl
73 & $-$23.04 & $-$17.49 & 28.93 & 24.02 & 19.42 & 4.75 &  \nl
74 & $-$28.87 & 5.74 & 29.44 & 25.02 & $>$20.67 & $<$4.03 & 0.923 \nl
75 & 21.03 & $-$20.72 & 29.52 & 24.75 & 19.86 & 5.02 &  \nl
76 & $-$28.87 & $-$7.40 & 29.81 & 23.88 & 19.23 & 4.71 &  \nl
77 & 3.39 & 29.81 & 30.00 & 25.39 & 19.75 & 5.68 &  \nl
78 & 20.44 & 21.96 & 30.00 & 22.94 & 19.59 & 3.28 &  \nl
79 & $-$8.05 & 28.93 & 30.03 & 24.32 & 19.27 & 5.40 &  \nl
80 & 30.34 & 2.07 & 30.41 & $>$26.43 & 20.90 & $>$5.52 &  \nl
81 & $-$30.42 & $-$0.64 & 30.43 & 25.25 & $>$20.54 & $<$4.31 &  \nl
82 & 25.47 & 17.60 & 30.96 & 21.62 & 17.27 & 4.38 & 0.261 \nl
83 & $-$20.60 & 23.50 & 31.25 & 23.09 & 19.48 & 3.35 &  \nl
84 & $-$30.51 & 7.59 & 31.44 & 25.84 & 20.84 & $<$4.95 &  \nl
85 & $-$7.89 & 30.68 & 31.67 & 23.95 & 20.71 & 3.50 & 0.709 \nl
86 & $-$20.61 & $-$24.10 & 31.71 & 25.20 & $>$20.64 & $<$4.34 &  \nl
87 & $-$25.08 & $-$19.84 & 31.98 & 23.81 & $>$20.41 & $<$3.18 &  \nl
88 & 6.89 & $-$31.63 & 32.37 & 23.01 & $>$19.87 & $<$2.84 &  \nl
89 & $-$5.70 & 32.45 & 32.95 & 25.13 & $>$20.73 & $<$4.10 &  \nl
90 & 21.00 & $-$25.55 & 33.07 & 23.17 & $>$19.94 & 3.28 &  \nl
91 & 12.70 & $-$31.23 & 33.71 & 23.09 & 19.74 & 3.30 & 0.280 \nl
92 & 12.01 & 31.51 & 33.72 & 24.98 & 18.84 & 6.17 &  \nl
93 & 30.44 & $-$15.29 & 34.06 & 25.46 & $>$20.63 & $<$4.59 &  \nl
94 & $-$34.31 & 3.22 & 34.46 & 22.37 & 18.19 & 4.60 & 1.011 \nl
95 & $-$29.84 & $-$17.76 & 34.73 & 21.94 & $>$19.67 & 1.99 & star \nl
96 & 34.15 & $-$7.14 & 34.89 & 24.32 & $>$20.31 & $<$3.58 &  \nl
97 & 33.89 & $-$9.64 & 35.23 & 25.36 & $>$21.08 & $<$4.25 &  \nl
98 & 21.72 & 28.07 & 35.49 & 21.04 & 17.70 & 3.32 &  \nl
99 & $-$4.22 & 35.62 & 35.87 & 24.93 & $>$20.44 & $<$4.23 &  \nl
100 & 20.42 & $-$30.08 & 36.35 & 25.43 & $>$20.74 & $<$4.56 &  \nl
101 & $-$36.70 & 1.18 & 36.72 & 25.11 & $>$20.72 & $<$4.04 &  \nl
102 & $-$16.26 & 33.54 & 37.27 & 22.88 & 17.48 & 5.40 &  \nl
103 & $-$37.32 & 4.74 & 37.61 & 22.80 & 19.78 & 3.44 &  \nl
104 & 28.72 & 24.42 & 37.70 & 25.20 & $>$20.64 & $<$4.32 &  \nl
105 & 35.80 & $-$12.86 & 38.04 & 20.58 & 17.35 & 3.23 &  \nl
106 & $-$17.98 & 33.67 & 38.17 & 26.05 & 20.87 & 5.19 &  \nl
107 & $-$36.32 & 13.26 & 38.67 & 24.71 & $>$20.50 & $<$3.90 &  \nl
108 & $-$36.46 & $-$12.89 & 38.67 & 25.18 & $>$20.50 & $<$4.60 &  \nl
109 & $-$27.55 & 27.28 & 38.77 & $>$25.65 & 18.92 & 6.70 &  \nl
110 & 34.97 & $-$17.17 & 38.96 & 24.95 & $>$20.33 & 4.50 &  \nl
111 & 8.85 & 38.13 & 39.15 & 25.35 & $>$20.53 & 4.44 &  \nl
112 & 10.54 & $-$38.00 & 39.43 & 23.88 & 19.60 & 3.95 &  \nl
113 & 25.69 & $-$30.06 & 39.55 & 24.89 & $>$20.69 & $<$4.10 &  \nl
114 & 33.93 & 20.38 & 39.58 & 19.32 & 15.76 & 3.54 &  \nl
115 & $-$19.96 & $-$34.58 & 39.93 & 23.25 & 18.73 & 4.52 &  \nl
116 & $-$12.78 & $-$38.04 & 40.13 & 26.05 & 20.64 & 5.38 &  \nl
117 & $-$19.76 & 35.25 & 40.41 & 24.82 & 20.08 & 4.70 &  \nl
118 & 40.99 & 2.89 & 41.09 & 25.07 & $>$20.42 & $<$4.27 &  \nl
119 & 6.02 & 40.80 & 41.24 & 25.41 & $>$20.91 & $<$4.22 &  \nl
120 & $-$24.14 & 33.73 & 41.48 & 23.91 & 18.12 & 5.61 &  \nl
121 & 13.87 & 39.14 & 41.53 & 18.47 & 16.51 & 1.98 &  \nl
122 & 39.05 & 15.44 & 41.99 & 25.03 & $>$20.57 & $<$4.15 &  \nl
123 & $-$27.47 & 32.93 & 42.88 & 21.43 & 17.72 & 3.55 &  \nl
124 & 42.50 & $-$6.50 & 42.99 & 25.47 & $>$20.50 & $<$4.68 &  \nl
125 & 30.50 & $-$30.38 & 43.05 & 23.48 & 19.62 & 3.75 &  \nl
126 & 30.05 & 32.11 & 43.97 & 23.60 & $>$20.39 & $<$3.17 &  \nl
127 & 2.40 & $-$43.93 & 44.00 & 23.40 & $>$20.65 & 3.54 &  \nl
128 & $-$7.62 & $-$43.35 & 44.01 & 24.02 & $>$19.93 & 3.92 & 0.892 \nl
129 & $-$43.33 & $-$12.68 & 45.15 & 21.40 & 18.81 & 2.65 &  \nl
130 & $-$2.61 & 45.10 & 45.17 & 22.78 & 19.68 & 2.81 & 0.515 \nl
131 & 45.22 & $-$0.65 & 45.22 & 23.80 & 19.70 & 3.68 &  \nl
132 & $-$22.36 & 39.83 & 45.68 & 22.96 & 18.11 & 4.80 &  \nl
133 & $-$45.80 & $-$3.69 & 45.95 & 24.99 & $>$20.66 & $<$4.20 &  \nl
134 & $-$45.39 & 7.71 & 46.04 & 24.27 & 21.29 & 3.84 & 0.668 \nl
135 & $-$29.58 & $-$35.71 & 46.37 & 23.75 & $>$20.00 & $<$3.36 &  \nl
136 & 41.71 & $-$20.37 & 46.41 & 23.15 & 19.16 & 4.16 &  \nl
137 & $-$18.80 & $-$42.53 & 46.50 & 22.16 & 17.75 & 4.37 &  \nl
138 & $-$44.23 & $-$15.52 & 46.87 & 25.49 & $>$21.33 & $<$3.87 &  \nl
139 & 43.40 & $-$18.17 & 47.05 & 23.34 & 19.60 & 3.90 &  \nl
140 & 25.16 & 39.78 & 47.07 & 25.08 & $>$20.50 & $<$4.34 &  \nl
141 & 39.97 & $-$25.18 & 47.24 & 25.30 & $>$20.73 & $<$4.03 &  \nl
142 & $-$46.76 & 7.72 & 47.40 & 23.76 & $>$21.38 & $<$3.21 & 0.638 \nl
143 & $-$35.86 & 31.33 & 47.62 & 23.63 & 17.97 & 5.62 &  \nl
144 & $-$42.94 & $-$20.97 & 47.79 & 23.68 & 19.44 & 4.06 &  \nl
145 & 6.81 & 47.53 & 48.01 & 24.92 & $>$21.24 & $<$4.49 &  \nl
146 & $-$23.20 & 42.20 & 48.16 & 23.66 & 18.50 & 5.10 &  \nl
147 & 7.56 & $-$47.91 & 48.50 & 25.27 & $>$20.37 & $<$4.86 &  \nl
148 & $-$11.74 & $-$47.45 & 48.88 & 24.82 & 20.08 & 4.51 &  \nl
149 & $-$17.58 & 45.75 & 49.01 & 21.89 & 17.24 & 4.65 &  \nl
150 & $-$40.93 & 27.08 & 49.08 & 23.89 & 20.92 & $<$3.06 &  \nl
151 & $-$48.20 & $-$9.91 & 49.20 & 24.17 & $>$20.39 & $<$3.48 &  \nl
152 & 37.93 & $-$31.39 & 49.23 & 25.05 & $>$20.52 & $<$4.54 &  \nl
153 & $-$38.48 & $-$30.80 & 49.29 & 22.62 & 18.17 & 4.50 &  \nl
154 & 48.25 & 10.73 & 49.43 & $>$26.51 & 20.92 & $>$5.50 &  \nl
155 & $-$24.98 & 42.78 & 49.54 & 22.45 & 17.53 & 4.85 &  \nl
156 & $-$32.30 & 37.83 & 49.74 & 25.19 & $>$20.53 & $<$4.36 &  \nl
\enddata
\tablenotetext{a}{Only objects within 50\arcs\ of the QSO are included}
\end{planotable}
\begin{planotable}{rrrrrrlr}
\footnotesize
\tablewidth{0pc}
\tablecaption{Absorption Lines in the FOS Spectrum of 3C~336}
\tablehead{
\colhead{No.} & \colhead{$\lambda_{\rm obs}$} & \colhead{$\sigma(\lambda)$} 
& \colhead{W$_{\rm obs}$} &
\colhead{$\sigma$(W)} & \colhead{S/N} & 
\colhead{ID} & \colhead{z$_{\rm abs}$} }
\startdata
1 & 1755.63 & 0.25 & 2.07 & 0.46 & 2.3 & SiII(1193) & 0.4713 \nl
  &         &      &      &      &     & CIII(977) & 0.7969  \nl
2 & 1759.39 & 0.31 & 2.80 & 0.59 & 2.2 & CII(1334) & 0.3184 \nl
3 & 1773.83 & 0.26 & 1.85 & 0.40 & 2.8 &  &  \nl
4 & 1795.02 & 0.25 & 2.14 & 0.40 & 3.0 &  &  \nl
5 & 1837.83 & 0.27 & 2.50 & 0.41 & 3.3 & SiIV(1393) & 0.3186 \nl
  &         &      &      &      &     & HI(972) & 0.8897  \nl
6 & 1842.47 & 0.26 & 2.26 & 0.39 & 3.2 & HI(1025) & 0.7963 \nl
7 & 1847.13 & 0.24 & 1.80 & 0.34 & 3.3 & CIII(977) & 0.8906 \nl
8 & 1854.04 & 0.27 & 1.49 & 0.31 & 3.9 & AlIII(1854) & -0.0004 \nl
  &         &      &      &      &     & SiII(1260) & 0.4710 \nl
  &         &      &      &      &     & OVI(1031) & 0.7966  \nl
9 & 1863.15 & 0.31 & 1.62 & 0.34 & 3.8 & AlIII(1862) & 0.0002 \nl
  &         &      &      &      &     & CII(1036) & 0.7978 \nl
  &         &      &      &      &     & OVI(1037) & 0.7956  \nl
10 & 1870.24 & 0.29 & 1.84 & 0.35 & 3.7 &  &  \nl
11 & 1939.14 & 0.25 & 1.78 & 0.27 & 4.9 & HI(1025) & 0.8905 \nl
12 & 1959.32 & 0.22 & 1.30 & 0.23 & 4.6 & CII(1036) & 0.8906 \nl
13 & 1963.25 & 0.25 & 1.51 & 0.25 & 5.0 & CII(1334) & 0.4711 \nl
14 & 1970.87 & 0.27 & 1.16 & 0.21 & 5.8 & SiII(1190) & 0.6556 \nl
15 & 1975.65 & 0.20 & 1.01 & 0.18 & 5.6 & SiII(1193) & 0.6556 \nl
16 & 1988.76 & 0.19 & 0.77 & 0.13 & 7.1 &  &  \nl
17 & 1997.34 & 0.22 & 1.05 & 0.17 & 6.9 & SiIII(1206) & 0.6555 \nl
18 & 2013.63 & 0.36 & 12.84 & 0.63 & 4.6 & HI(1215) & 0.6564 \nl
19 & 2039.61 & 0.21 & 1.03 & 0.20 & 5.0 & CIV(1548) & 0.3174 \nl
20 & 2042.93 & 0.22 & 0.81 & 0.17 & 6.0 & CIV(1550) & 0.3174 \nl
21 & 2050.43 & 0.21 & 1.98 & 0.24 & 5.1 & SiIV(1393) & 0.4711 \nl
   &         &      &      &      &     & NV(1238) & 0.6551 \nl
22 & 2069.03 & 0.21 & 0.81 & 0.17 & 5.6 &  &  \nl
23 & 2083.62 & 0.22 & 0.74 & 0.16 & 6.5 &  &  \nl
24 & 2086.53 & 0.22 & 1.33 & 0.20 & 6.1 & SiII(1260) & 0.6554 \nl
25 & 2103.13 & 0.24 & 1.56 & 0.21 & 6.3 &  &  \nl
26 & 2155.96 & 0.24 & 0.78 & 0.16 & 6.7 & OI(1302) & 0.6557 \nl
27 & 2159.20 & 0.29 & 0.65 & 0.16 & 7.0 & SiII(1304) & 0.6554 \nl
28 & 2167.40 & 0.23 & 1.48 & 0.20 & 6.5 & SiIII(1206) & 0.7964 \nl
29 & 2183.84 & 0.19 & 2.46 & 0.24 & 5.4 & HI(1215) & 0.7964 \nl
30 & 2209.82 & 0.18 & 0.95 & 0.14 & 6.7 & CII(1334) & 0.6559 \nl
31 & 2221.41 & 0.19 & 1.94 & 0.20 & 6.3 &  &  \nl
32 & 2250.67 & 0.17 & 1.16 & 0.14 & 7.3 & SiII(1190) & 0.8907 \nl
33 & 2253.91 & 0.18 & 0.64 & 0.11 & 8.0 &  &  \nl
34 & 2256.29 & 0.18 & 1.20 & 0.14 & 7.5 & SiII(1193) & 0.8908 \nl
35 & 2278.33 & 0.19 & 0.76 & 0.11 & 8.8 & CIV(1548) & 0.4716 \nl
36 & 2281.42 & 0.16 & 1.50 & 0.15 & 6.9 & CIV(1550) & 0.4712 \nl
   &         &      &      &      &     & SiIII(1206) & 0.8909 \nl
37 & 2298.73 & 0.18 & 5.13 & 0.28 & 6.8 & HI(1215)  & 0.8909 \nl
38 & 2308.26 & 0.27 & 0.50 & 0.11 & 11.7 & SiIV(1393) & 0.6561 \nl
39 & 2344.12 & 0.22 & 0.73 & 0.06 & 23.2 & FeII(2344) & 0.0000 \nl
   &         &      &      &      &      & SiII(1304) & 0.7971 \nl
   &         &      &      &      &      & NV(1238) & 0.8922  \nl
40 & 2374.17 & 0.29 & 0.78 & 0.11 & 14.6 & FeII(2374) & 0.0002 \nl
41 & 2383.17 & 0.25 & 2.54 & 0.16 & 13.3 & FeII(2382) & 0.0002 \nl
   &         &      &      &      &      & SiII(1808) & 0.3181 \nl
   &         &      &      &      &      & SiII(1260) & 0.8908 \nl
42 & 2398.25 & 0.30 & 0.75 & 0.11 & 14.4 & CII(1334) & 0.7971 \nl
43 & 2463.09 & 0.24 & 0.79 & 0.12 & 10.7 & OI(1302) & 0.8915 \nl
44 & 2466.72 & 0.26 & 0.94 & 0.13 & 10.9 & SiII(1304) & 0.8911 \nl
45 & 2504.44 & 0.26 & 0.93 & 0.13 & 10.7 & SiIV(1393) & 0.7969 \nl
46 & 2523.26 & 0.25 & 2.15 & 0.18 & 10.5 & CII(1334) & 0.8907 \nl
47 & 2528.41 & 0.25 & 1.06 & 0.13 & 11.1 & SiII(1526) & 0.6561 \nl
48 & 2564.08 & 0.30 & 0.53 & 0.11 & 11.8 & CIV(1548) & 0.6562 \nl
49 & 2567.90 & 0.32 & 0.58 & 0.12 & 11.9 & CIV(1550) & 0.6559 \nl
50 & 2586.75 & 0.28 & 0.63 & 0.12 & 11.1 & FeII(2586) & 0.0000 \nl
51 & 2600.11 & 0.26 & 0.75 & 0.12 & 11.4 & FeII(2600) & 0.0000 \nl
52 & 2636.05 & 0.26 & 1.51 & 0.15 & 11.1 & SiIV(1393) & 0.8913 \nl
53 & 2652.86 & 0.27 & 1.05 & 0.13 & 12.2 & SiIV(1402) & 0.8912 \nl
54 & 2664.02 & 0.37 & 0.99 & 0.15 & 12.9 & FeII(1608) & 0.6563 \nl
55 & 2767.18 & 0.30 & 1.00 & 0.14 & 12.3 & AlII(1670) & 0.6562 \nl
56 & 2782.40 & 0.24 & 2.48 & 0.17 & 10.2 & CIV(1548) & 0.7972 \nl
57 & 2787.21 & 0.23 & 1.79 & 0.15 & 10.2 & CIV(1550) & 0.7973 \nl
58 & 2796.08 & 0.27 & 0.96 & 0.13 & 11.6 & MgII(2796) & -0.0001 \nl
59 & 2803.69 & 0.26 & 1.22 & 0.14 & 11.6 & MgII(2803) & 0.0001 \nl
60 & 2853.42 & 0.36 & 0.81 & 0.14 & 12.8 & MgI(2852) & 0.0002 \nl
61 & 2887.50 & 0.31 & 1.32 & 0.16 & 11.3 & SiII(1526) & 0.8913 \nl
62 & 2927.91 & 0.23 & 1.73 & 0.13 & 13.3 & CIV(1548) & 0.8912 \nl
63 & 2933.01 & 0.24 & 1.73 & 0.13 & 14.1 & CIV(1550) & 0.8913 \nl
64 & 3042.13 & 0.28 & 0.46 & 0.08 & 16.0 & FeII(1608) & 0.8913 \nl
65 & 3160.42 & 0.24 & 1.00 & 0.11 & 13.0 & AlII(1670) & 0.8915 \nl
\enddata
\end{planotable}
\begin{planotable}{rrrrrrlr}
\footnotesize
\tablewidth{0pc}
\tablecaption{Absorption Lines in the KPNO Spectrum of 3C~336}
\tablehead{
\colhead{No.} & \colhead{$\lambda_{\rm obs}$} & \colhead{$\sigma(\lambda)$} 
& \colhead{W$_{\rm obs}$} &
\colhead{$\sigma$(W)} & \colhead{S/N} & 
\colhead{ID} & \colhead{z$_{\rm abs}$} }
\startdata
  1  &  3506.60  &  0.05  &  0.96  &  0.06 &   30 &   Fe II (2382)  &   0.4717\nl
     &           &        &        &       &      &   Al III (1854) &   0.8906\nl
  2  &  3522.18  &  0.05  &  0.36  &  0.06 &   30 &   Al III (1862) &   0.8908\nl
  3  &  3683.94  &  0.06  &  0.66  &  0.04 &   50 &   Mg II (2796)  &   0.3174\nl
  4  &  3693.09  &  0.06  &  0.57  &  0.04 &   50 &   Mg II (2803)  &   0.3173\nl
  5  &  3806.37  &  0.04  &  0.21  &  0.03 &   60 &   Fe II (2586)  &   0.4715\nl
  6  &  3825.62  &  0.06  &  0.74  &  0.03 &   60 &   Mg II (2796)  &   0.3681\nl
     &           &        &        &       &      &   Fe II (2600)  &   0.4713\nl
  7  &  3833.92  &  0.05  &  0.40  &  0.03 &   60 &   Mg II (2803)  &   0.3675\nl
  8  &  3881.31  &  0.03  &  1.34  &  0.03 &   60 &   Fe II (2344)  &   0.6557\nl
  9  &  3931.53  &  0.06  &  0.84  &  0.03 &   70 &   Fe II (2374)  &   0.6558\nl
 10  &  3933.56  &  0.06  &  0.30  &  0.03 &   70 &   Ca II (3933)  &   0.0000\nl
 11  &  3945.13  &  0.04  &  1.74  &  0.03 &   70 &   Fe II (2382)  &   0.6557\nl
 12  &  3968.43  &  0.05  &  0.20  &  0.03 &   70 &   Ca II (3968)  &   0.0000\nl
 13  &  4115.17  &  0.05  &  1.26  &  0.03 &   70 &   Mg II (2796)  &   0.4716\nl
 14  &  4125.71  &  0.03  &  1.04  &  0.03 &   70 &   Mg II (2803)  &   0.4716\nl
 15  &  4198.10  &  0.05  &  0.13  &  0.03 &   70 &   Mg I (2852)   &   0.4715\nl
 16  &  4282.71  &  0.03  &  1.38  &  0.03 &   70 &   Fe II (2586)  &   0.6557\nl
 17  &  4305.15  &  0.04  &  1.84  &  0.03 &   70 &   Fe II (2600)  &   0.6557\nl
 18  &  4432.53  &  0.03  &  1.46  &  0.03 &   60 &   Fe II (2344)  &   0.8908\nl
 19  &  4489.63  &  0.05  &  0.86  &  0.05 &   40 &   Fe II (2374)  &   0.8908\nl
 20  &  4505.26  &  0.05  &  1.86  &  0.05 &   40 &   Fe II (2382)  &   0.8908\nl
\enddata
\end{planotable}
\begin{planotable}{rrrrrrlr}
\footnotesize
\tablewidth{0pc}
\tablecaption{Absorption Lines in the Keck Spectrum of 3C~336}
\tablehead{
\colhead{No.} & \colhead{$\lambda_{\rm obs}$} & \colhead{$\sigma(\lambda)$} 
& \colhead{W$_{\rm obs}$} &
\colhead{$\sigma$(W)} & \colhead{S/N} & 
\colhead{ID} & \colhead{z$_{\rm abs}$} }
\startdata
1 & 4919.09 & 0.08 & 2.00 & 0.03 & 108.5 & FeII(2600) & 0.8918 \nl
2 & 5026.10 & 0.10 & 0.76 & 0.03 & 127.4 & MgII(2796) & 0.7974 \nl
3 & 5038.61 & 0.16 & 0.55 & 0.03 & 121.9 & MgII(2803) & 0.7972 \nl
4 & 5288.89 & 0.07 & 2.97 & 0.04 & 92.4 & MgII(2796) & 0.8914 \nl
5 & 5302.19 & 0.07 & 2.28 & 0.04 & 107.4 & MgII(2803) & 0.8913 \nl
6 & 5396.52 & 0.08 & 0.53 & 0.02 & 193.9 & MgI(2852) & 0.8915 \nl
7 & 5891.10 & 0.09 & 0.37 & 0.02 & 178.8 & NaI(5891) & 0.0000 \nl
8 & 5897.36 & 0.13 & 0.34 & 0.02 & 173.4 & NaI(5897) & 0.0000 \nl
9 & 6516.21 & 0.15 & 0.27 & 0.02 & 173.1 & CaII(3934) & 0.6561 \nl
\enddata
\end{planotable}
\newpage
\begin{planotable}{rccccrrr}
\footnotesize
\tablewidth{0pc}
\tablecaption{Absorption Strength and Galaxies $<100h^{-1}$ kpc from Line of Sight\tablenotemark{a}}
\tablehead{
\colhead{ID No.} & \colhead{Redshift} & \colhead{$L/L^{\ast}_B$} 
& \colhead{$L/L^{\ast}_K$} & \colhead{$D$} & \colhead{W$_{\lambda}$(MgII)} &
\colhead{W$_{\lambda}$(CIV)} & \colhead{W$_{\lambda}$(Ly $\alpha$)}  \nl
\colhead{} & \colhead{} & \colhead{(rest)} & 
\colhead{(rest)} & \colhead{($h^{-1}$ kpc)} & \colhead{(\AA)} 
& \colhead{(\AA)} & \colhead{(\AA)}
 }
\startdata
82 & 0.261 & 0.10 & 0.25 & 83.0 & $<$0.24 & $<$0.99 & \nodata \nl
91 & 0.280 & 0.04 & 0.03 & 94.8 & $<$0.20 & $<$0.58 & \nodata \nl
16 & 0.318 & 0.58 & 0.99 & 36.1 & 0.50 & 0.30 & \nodata \nl
45 & 0.368 & 1.28 & 1.78 & 74.4 & [0.29]\tablenotemark{e} & $<$0.54 & \nodata \nl
52 & 0.368 & 0.04 & 0.06 & 82.6 & \nodata & \nodata & \nodata \nl
4 & 0.472 & 0.18 & 0.10 & 21.7 & 0.86 & 0.52 & $<$1.40 \nl
12 & 0.565 & 0.07 & 0.03 & 40.9 & $<$0.04\tablenotemark{c}& $<$0.35 & $<$0.80 \nl
10 & 0.635 & 0.11 & 0.02 & 41.7 & $<$0.03\tablenotemark{c} & $<$0.30 & 0.47 \nl
26\tablenotemark{b} & 0.656 & 0.24 & 0.07 & 64.7 & 1.68 & 0.33 & 7.75 \nl
30 & 0.702 & 1.53 & 0.97 & 73.4 & 0.04\tablenotemark{c} & $<$0.29 & 0.48 \nl
11 & 0.798 & 1.06 & 0.82 & 46.5 & 0.38 & 1.38 & 1.37 \nl
41 & 0.828 & 0.38 & 0.47 & 93.1 & $<$0.08 & $<$0.30 & 1.06\tablenotemark{d} \nl
3 & 0.892 & 0.94 & 0.39 & 15.0 & 1.57 & 0.91 & 2.71 \nl
\enddata
\tablenotetext{a}{Assuming $H_0=50$ \kms Mpc$^{-1}$ and $q_0=0.05$ for
luminosities; distances $D$ are given in units of $h_{100}^{-1}$ kpc. Only galaxies
having redshifts placing them in the foreground relative to 3C~336 are included.}
\tablenotetext{b}{Assuming that object \#26 is responsible for the damped Lyman $\alpha$ absorption,
which we consider unlikely (see text).}
\tablenotetext{c}{Measurement from a Keck/HIRES spectrum of 3C~336 (Churchill 1996,
private communication)}
\tablenotetext{d}{It is very uncertain whether this spectral feature is a Lyman $\alpha$
line at $z=0.828$}
\tablenotetext{e}{Line strength is uncertain because it is blended with a line from another
absorption system}
\end{planotable}


\end{document}